\documentclass{article}

\usepackage{PRIMEarxiv}

\usepackage[utf8]{inputenc} 
\usepackage[T1]{fontenc}    
\usepackage{hyperref}       
\usepackage{url}            
\usepackage{booktabs}       
\usepackage{amsfonts}       
\usepackage{nicefrac}       
\usepackage{microtype}      
\usepackage{lipsum}
\usepackage{fancyhdr}       
\usepackage{graphicx}       
\graphicspath{{media/}}     
\usepackage[round]{natbib}
\usepackage{setspace}

\title{Matter \& Mind Matter
\thanks{\textit{\underline{Citation}}: 
\textbf{Authors. Title. Pages.... DOI:000000/11111.}} 
}

\author{
  Tom Birkoben \\
  Kiel University \\
  Faculty of Engineering \\
  Institute for Electrical Engineering and Information Engineering\\
  Chair of Nanoelectronics\\
  \texttt{tobi@tf.uni-kiel.de} \\
   \And
  Hermann Kohlstedt 
  \thanks{Corresponding author} \\
  Kiel University \\
  Faculty of Engineering \\
  Institute for Electrical Engineering and Information Engineering\\
  Chair of Nanoelectronics\\
  \texttt{hko@tf.uni-kiel.de} \\
}

\begin{document}
\maketitle

\begin{abstract}
\setstretch{1.0}
As a result of a hundred million years of evolution, living animals have adapted extremely well to their ecological niche. Such adaptation implies species-specific interactions with their immediate environment by processing sensory cues and responding with appropriate behavior. Understanding how living creatures perform pattern recognition and cognitive tasks is of particular importance for computing architectures: by studying these information pathways refined over eons of evolution, researchers may be able to streamline the process of developing more highly advanced, energy efficient autonomous systems. 

With the advent of novel electronic and ionic components along with a deeper understanding of information pathways in living species, a plethora of opportunities to develop completely novel information processing avenues are within reach.

Here, we describe the basal information pathways in nervous systems, from the local neuron level to the entire nervous system network. The dual importance of local learning rules is addressed, from spike timing dependent plasticity at the neuron level to the interwoven morphological and dynamical mechanisms of the global network. Basal biological principles are highlighted, including phylogenies, ontogenesis, and homeostasis, with particular emphasis on network topology and dynamics. While in machine learning system training is performed on virgin networks without any a priori knowledge, the approach proposed here distinguishes itself unambiguously by employing growth mechanisms as a guideline to design novel computing architectures. Including fundamental biological information pathways that explore the spatiotemporal fundamentals of nervous systems has untapped potential for the development of entirely novel information processing systems. Finally, a benchmark for neuromorphic systems is suggested. 

\end{abstract}

\keywords{Bio-inspired Computing \and Phylogenesis \and Ontogenesis \and Homeostasis \and Artificial spatio-temporal networks}

\section{Introduction}
Is it truly possible to implement higher brain functions, such as perception or consciousness, in engineered systems? This question has been frequently raised over the last few decades and has led to distinct views over time, as both neurobiological understanding and available computational capabilities advanced \citep{churchlandCouldMachineThink1990, churchland1999densmore, aleksanderHowBuildMind2001, hawkinsIntelligence2004, kochCanMaschinesBe2008, dehaeneConsciousnessBrainDeciphering2014, dehaeneWhatConsciousnessCould2017}. The essence of this question goes back to the fundamental relation between matter and mind, which was addressed as early as ancient Greece, and emerged in the principle of “Dualism” most famously defended by the philosopher René Descartes in the sixteen century\citep{ostenfeldAncientGreekPsychology2018}. Descartes postulated that the body (matter) and the mind are distinct and separate units in human beings because he could not imagine that mental phenomena could be explained by natural mechanisms \citep{damasioDescartesErrorEmotion2004}.  However, the invention of electroencephalography (EEG) and imaging techniques, such as functional magnetic resonance imaging (fMRI), enabled the study of inner information processing in the human brain and individuals’ states of consciousness \citep{varelaBrainwebPhaseSynchronization2001, noirhommeConsciousnessUnconsciousnessEEG2014, kriegeskorteCognitiveComputationalNeuroscience2018,demertziHumanConsciousnessSupported2019}. As a result, the strict distinction between matter and mind has become blurry \citep{stormConsciousnessRegainedDisentangling2017, dehaeneCognitiveNeuroscienceConsciousness2001}. Strong evidence has been found that the inner representation of the human brain (the mind) is related to its neurochemistry (the matter), e.g. the amount and type of neurotransmitters and/or drugs within the nervous system\citep{tagliazucchienzoLargescaleSignaturesUnconsciousness2016, perryAcetylcholineMindNeurotransmitter1999}. It is therefore worthwhile to reconsider the relationship between mind and matter when engineering artificial systems to exhibit higher brain functions by considering recent progress in nanoelectronics and neurobiology.  

This perspective on future computing is motivated by three key aspects. First by the recent, growing movement to reboot the entire field of computing, i.e. how data are processed. Second by state-of-the-art, fundamental progress in neurosciences, including the fields of complex networks and dynamic brain states. Third by advances in materials science and nanoelectronics that have led to, e.g., memristive devices, nanoparticle/nanowire networks, and fluidic memristors, providing new functionality in electronics, such as synaptic-like plasticity or spatio-temporal networks \citep{xiaMemristiveCrossbarArrays2019, bianStimuliResponsiveMemristive2021, mallinsonAvalanchesCriticalitySelforganized2019, robinModelingEmergentMemory2021, kuncic2021neuromorphic}. With the foreseen restrictions on current digital computing, the question “What comes next?” finds its answer in merging novel discoveries made on the nervous system’s information pathways with the development of novel electronic devices, paving the way to an entirely new kind of computing.

In this perspective, we present a concept of an artificial spatio-temporal network which uses temporal and structural mechanisms in nervous systems as guidelines. It addresses the important, interwoven spatiotemporal aspects of information pathways and processing in nervous systems \citep{beggsNeuronalAvalanchesNeocortical2003, bullmoreComplexBrainNetworks2009,chialvoEmergentComplexNeural2010, spornsNetworksBrain2011, fornitoFundamentalsBrainNetwork2016, bassettNetworkNeuroscience2017}. The state of nervous system criticality combined with the blooming and pruning of nervous cells during growth might be an interesting guideline to develop new computing principles \citep{beggsNeuronalAvalanchesNeocortical2003, kaiserSpatialGrowthRealworld2004, kinouchiOptimalDynamicalRange2006, chialvoEmergentComplexNeural2010, kaiserHierarchyDynamicsNeural2010, huttmarc-thorstenPerspectiveNetworkguidedPattern2014, kaiserMechanismsConnectomeDevelopment2017, agiNeuronalStrategiesMeeting2020}. Components essential for artificial spatio-temporal networks and a pathway to realize it, are presented, including biological fundamentals such as phylogenies, ontogenesis, and homeostasis \citep{lvtrupPhylogenesisOntogenesisEvolution1987, tordayHomeostasisMechanismEvolution2015}. However, these basal biological mechanisms alone might not be sufficient to establish mental functions in artificial systems; we therefore include the temporal binding hypothesis developed in neuroscience as a further essential guideline\citep{singerVisualFeatureIntegration1995, engelDynamicPredictionsOscillations2001, varelaBrainwebPhaseSynchronization2001, uhlhaasNeuralSynchronyCortical2009, sheffieldBindingSolution2015}. The synchronized firing of neural ensembles across different brain regions is treated as a fundamental neural mechanism that defines how a hierarchical network structure, such as the brain, can integrate several sensory inputs to determine the unity of an object (for example linking form, color, size, motion, etc., together) \citep{buzsakiRhythmsBrain2006,schechterHowBrainGets1996}. Opportunities and possible limitations of this approach towards implementing higher brain functions in artificial systems, such as perception and consciousness, will be discussed. 
The paper is arranged as follows:

In Chapter 2, the current status of computing architectures is summarized. In Chapter 3, we present a condensed overview of advanced device components, with a focus on memristive switching devices. We subsequently address spatiotemporal information processing in nervous systems, including network structure, network dynamics, and homeostasis in Chapter 4. An artificial spatio-temporal network concept is introduced in Chapter 5, where we hypothesize on which information pathways might lead to higher brain functions in engineered systems based on hardware-oriented electrochemical electronics but also discuss current limitations of this approach. In Chapter 6 a possible benchmark is discussed for bio-inspired systems.  Chapter 7 provides a discussion on the practical implementation of an artificial spatio-temporal network, to mimic basal biological information pathways.

\section{The current state of information technology}
\label{sec:headings}

The sixties marked the beginning of a glorious time in information technology as the tremendous opportunities of silicon technology merged with the concept of Boolean computing, resulting in the first digital revolution \citep{berlinManMicrochipRobert2005}.

This development followed the exponential increase over time of electronic components integration on a chip predicted by Gordon E. Moore, combined with a sequential data processing architecture comprising a central processing unit and memory for data storage, for which Alain Turing and John von Neumann laid the foundation years before \citep{mooreCrammingMoreComponents1965, turingComputableNumbersApplication1937,j.vonneumannFirstDraftReport1993}. The tremendous technological and economical success of the digital revolution is still going strong today with seemingly no end in sight. Billions of transistors on a single processor chip, displaying features as small as about 10 nm in size and clock frequencies of a few GHz, are the current standard in CMOS (Complementary Metal Oxide Semiconductor) technology, representing the backbone of today’s semiconductor industry  \citep{j.m.veendrickNanometerCMOSICs2017, masuharaFutureLowPowerElectronics2016, kurinecEnergyEfficientComputing2019}. However, during the last couple of years, dark clouds have appeared on the horizon for the semiconductor industry. The envisioned goal of downscaling devices with every new circuit generation to the nm level has created an ongoing need to develop ever more sophisticated and expensive fabrication tools for e.g. lithography, dry-etching, and layer deposition \citep{dennardDesignIonimplantedMOSFET1974, hofflingerNewVistasNanoelectronics2016,radamsonStateArtFuture2020}. As a result, each new circuit generation entails an increasing economic risk for semiconductor companies. Moreover, over the last few decades, progress in processor core clock rates have overtaken memory access and access times, leading to a cumbersome situation where data transmission between the arithmetic logic unit (ALU) and memories dominates instead of the arithmetic information process itself. This system level-related challenge is called memory latency (or memory gap) and is a consequence of the von Neumann bottleneck, where data is processed sequentially \citep{backusCanProgrammingBe1978, iniewskiCMOSProcessorsMemories2010, j.m.veendrickNanometerCMOSICs2017}.  Two major obstacles restrict the further development of information technology, namely limitations in downscaling at the device level, and memory latency on the architecture. Although society is experiencing a second digital revolution via the resurgence of artificial intelligence (AI) and the Internet of Things (IOT), Moore’s law, which has been driving the computer industry for decades, is becoming outdated as the limits of device integration and/or economical boundaries have now been reached. The incredible advances made by the first digital revolution based on binary “0” and “1” computation combined with the latest achievements in the field of machine learning led to great progress in speech and pattern recognition, while rendering autonomous driving tangible. Yet, additional challenges are growing increasingly problematic behind the scenes. Huge, power consuming hardware systems in the form of cloud servers are now mandatory to support recent advancements in AI and the IOT. This is why global digital players, such as Google, Amazon, and Facebook, as well as bitcoin trading platforms need energy-hungry server farms \citep{jonesHowStopData2018}. 

On the system level, and in particular since the advent of the internet and the renewed interest in AI, the power consumption of the digital world is growing without limits, in increasing conflict with sustainable and climate-neutral resource management. Moreover, future autonomous electric vehicles require both high recognition capability and low power consumption. It therefore is hardly surprising that the semiconductor world is currently in an era of upheaval, turning a new page on information processing based on novel computing architectures and advanced hardware components.

\section{Advanced computing architectures and novel electronic devices}
\label{sec:advanced}

The aim of this section is to give a short survey on novel computing architectures and advanced electronic devices. We do not intend to present a comprehensive overview but instead to give a taste of the developments currently being pursued to overcome the limitations of digital computing and to establish new computing primitives. To simplify access to the different research areas for interested readers, we discuss seminal and overview papers and present recently published pioneering research. Nonetheless, we are aware that the given reference list is far from exhaustive. In addition, this section is critical to understanding the similarities, and most importantly the distinctions, between artificial spatio-temporal networks and standard neuromorphic computation presented in section 5.
While traditional von Neumann computing continues to dominate the ICT scene, recent groundbreaking innovations in alternative computing architectures and advanced electronic devices have become hard to ignore \citep{yangMemristiveDevicesComputing2012,schumanSurveyNeuromorphicComputing2017,burrNeuromorphicComputingUsing2017,merollaMillionSpikingneuronIntegrated2014,kendallBuildingBlocksBraininspired2020,kuncic2021neuromorphic}. 

These developments are threefold. Firstly, somewhat older concepts, such as artificial neural networks (ANN) leading to Deep Learning (DL) systems, have received an impressive performance boost through novel and efficient algorithms paired with more powerful electronics hardware \citep{lecunDeepLearning2015}. Secondly, new technologies, such as Quantum Computing and Reservoir Computing (RC), have appeared, leading to remarkable results \citep{merminQuantumComputerScience2007,aruteQuantumSupremacyUsing2019,gauthierNextGenerationReservoir2021}. Thirdly, in the field of nanoelectronics, a plethora of advanced device structures and novel functional components has led to a rethink of traditional computing architecture, paving the way to in-memory computing that circumvents the von-Neumann bottleneck \citep{burrNeuromorphicComputingUsing2017, ielminiInmemoryComputingResistive2018, kendallBuildingBlocksBraininspired2020, kasparRiseIntelligentMatter2021}. In Fig. 1 a shamrock-like illustration highlights these three research areas.

\begin{figure} [ht!]
	\centering
	\includegraphics[width=0.48\textwidth]{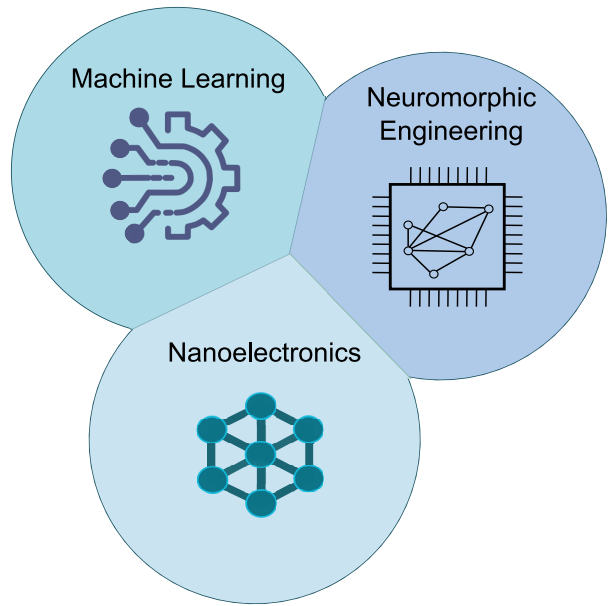}
	\caption{A shamrock-like illustration of the three development areas, which characterize the currently expansive development in the field of Artificial Intelligence (AI).}
	\label{fig:shamrock}
\end{figure}

The first leaf representing Machine Learning encompasses Artificial Neural Networks (ANN), Spiking Neural Networks (SNN), Reservoir Computing (RC), Long Short Term Memory (LSTM), and Deep Learning (DL) systems \citep{hintonDeepNeuralNetworks2012}.  The foundations of Neural Networks were laid by McCulloch and Pitts \citep{mccullochLOGICALCALCULUSIDEAS}, Rosenblatt´s Perceptron \citep{rosenblattPerceptronProbabilisticModel1958} for ANNs, and von Neumann´s postulate of SNNs in 1956. More recent inventions from Jäger (Reservoir Computing) \citep{jagerEchoStateApproach2001}, Hochreiter$\&$Schimdhuber (Long-/Short-Term Memory) \citep{hochreiterLongShortTermMemory1997}) and Hinton (DL) \citep{hintonDeepNeuralNetworks2012}) have advanced the field one huge step forward and comprise the backbone of today’s AI.

In the second leaf, the field of neuromorphic engineering, initiated by Carver Mead and Mohawa $\&$ Rodney Douglas, seeks to mimic the basal mechanisms of information processing in nervous systems via an essentially hardware-oriented approach \citep{meadAnalogVLSINeural1989, mahowaldSiliconNeuron1991, meadHowWeCreated2020}. In recent years, great progress has been made in the development of bio-inspired processors.  Here, event-based spiking neural networks (SNNs) in the form of either mixed (analog and digital) or strictly digital signal processing provides novel opportunities for low-power data processing \citep{indiveriNeuromorphicSiliconNeuron2011, merollaMillionSpikingneuronIntegrated2014, peiArtificialGeneralIntelligence2019, kendallBuildingBlocksBraininspired2020,frenkelBottomUpTopDownNeural2021}. Interestingly, some of the spiking neuromorphic circuits work at biologically relevant frequencies, exhibiting low energy consumption. One point of merit for neuromorphic engineering is their energy per synaptic operation (SOP), which is in the pJ to nJ range for neuroprocessors \citep{frenkelBottomUpTopDownNeural2021, kendallBuildingBlocksBraininspired2020, schumanSurveyNeuromorphicComputing2017}. 

Hence, the incorporation of relatively few basal mechanisms of biological information processing, such as leaky-integrated firing, axon delays, and local learning rules, can lead to significant improvements in resource management. 

Recent advances in the field of nanoelectronics devices, such as memristive devices, nanoparticle networks, nanowire networks, or memristive fluids, compose the third leaf of advanced computer architecture. Research in silicon nanoelectronics is dominated by the development of new field effect transistors (FET) \citep{karbalaeiSectorialSchemeGateallaround2021, radamsonStateArtFuture2020} for the next generation of CMOS circuits, as well as entirely novel devices and materials exhibiting advanced functionalities \citep{senguptaMagneticTunnelJunction2016,zhangOrganismicMaterialsNeumann2020,minnaiFacileFabricationComplex2017, lequeuxMagneticSynapseMultilevel2016, sangwanNeuromorphicNanoelectronicMaterials2020, sungPerspectiveReviewMemristive2018, kasparRiseIntelligentMatter2021}. In particular, the memristor (originating from memory and resistor, also called memristive device) is a two terminal device that exhibits attractive features for various applications in the post-Moore area, generating considerable interest. Memristive devices were intensively studied in the sixties and seventies \citep{hickmottLowFrequencyNegative1962, argallSwitchingPhenomenaTitanium1968, dearnaleyElectricalPhenomenaAmorphous1970}. The field was further propelled forward by the establishment of the theoretical background of memristors by Leon Chua (1971), with the corresponding experimental realization and interpretation by Hewlett-Packard (HP)-Labs (2008) \citep{l.chuaMemristorTheMissingCircuit1971, strukovMissingMemristorFound2008a}. Over the years, numerous books and reviews have covered fundamental and practical properties of memristive devices and their related circuits \citep{tetzlaffMemristorsMemristiveSystems2014,ielminiResistiveSwitchingFundamentals2016, sungPerspectiveReviewMemristive2018, xiaMemristiveCrossbarArrays2019, liMemristiveCrossbarArrays2021}. 

So far we have described nanoelectronic devices fabricated using top-down methods, where the layers are deposited on an entire wafer and the devices are patterned by lithography and dry-etching \citep{peaseChargeNotCharge2010, donnellyPlasmaEtchingYesterday2013, oluwatosinabegundeOverviewThinFilm2019}. In bottom-up approaches, functional materials are deposited or synthesized to obtain networks, such as irregular nanowires and/or 3D textures. Often the self-assembly capabilities of materials are exploited to create complex structures. Top-down and bottom-up approaches are habitually combined to create the electrical connections necessary to characterize the structures \citep{kronholzSelfAssemblyDiblockCopolymerMicelles2006}. In the context of bio-inspired computing, we would like to highlight here the work done on nanowire networks \citep{stiegEmergentCriticalityComplex2012, asayesh-ardakaniAtomicResolutionStudies2013, pantoneMemristiveNanowiresExhibit2018, hochstetterAvalanchesEdgeofchaosLearning2021, zhuInformationDynamicsNeuromorphic2021, loefflerTopologicalPropertiesNeuromorphic2020, mallinsonAvalanchesCriticalitySelforganized2019, pikeAtomicScaleDynamics2020}. The structure of such networks, and in particular their dynamic properties, reflect basal functionalities as observed in nervous systems, such as small-word connectivity and self-organized criticality (SOC) \citep{wattsCollectiveDynamicsSmallworld1998, beggsNeuronalAvalanchesNeocortical2003}. 
We would like to emphasize that the three ICT research areas shown in Fig. 1 are not independent from one another: there is considerable overlap between them, which has proven to be mutually beneficial.

\subsection{Advanced computing architectures}

Here we present a few concepts of novel and reconsidered computing architectures. We would like to emphasize that the following four examples were chosen to demonstrate the diversity of the field but are not intended to give a comprehensive overview. The icons in Fig. 2 represent different computing principles. 

\begin{figure} [ht!]
	\centering
	\includegraphics[width=0.8\textwidth]{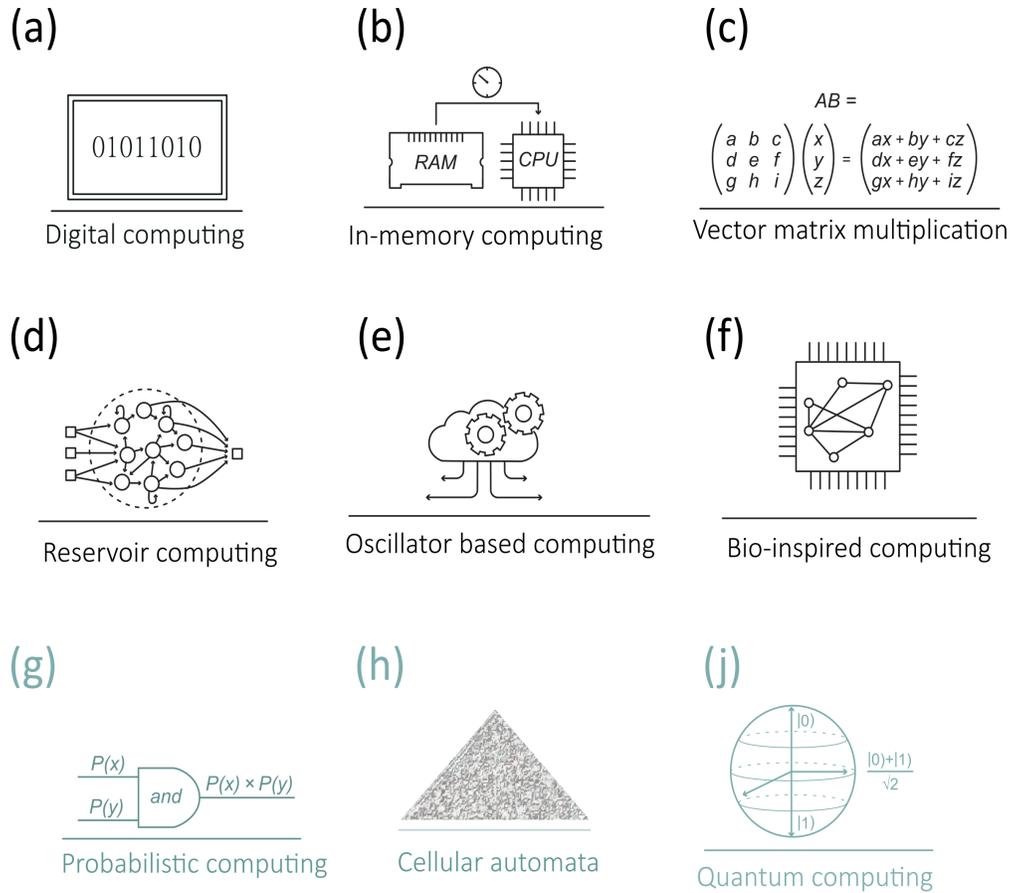}
	\caption{The illustration shows snapshots of different computing strategies. (a) digital computing, (b) in-memory computing, (c) matrix multiplication, (d) reservoir computing, (e) oscillatory computing, (f) bio-inspired computing, (g) probabilistic computing, (h) cellular automata, (i) quantum computing.}
	\label{fig:computing_strategies}
\end{figure}

For concepts other than those shown in Fig. 2, such as quantum computing, cellular automata, and probabilistic computing, we refer to the literature \citep{merminQuantumComputerScience2007, serbUnsupervisedLearningProbabilistic2016, baatarCellularNanoscaleEnsory2010, zhangOrganismicMaterialsNeumann2020, kariPrinciplesStochasticComputing}. We focus on comparing today’s digital computing to in-memory computing, vector matrix multiplication, reservoir computing, oscillatory computing, and bio-inspired computing (see icons in Fig. 2).

In order to overcome the von Neumann bottleneck of digital computing (Fig. 2(a)), near-memory computing was developed in 1990 \citep{d.pattersonCaseIntelligentRAM1997}. Here, the strict separation of an arithmetic logic unit communicating with several distinct memories was eliminated. Part of the computational tasks was performed within the memory itself, leading to more efficient computing. This development has recently shifted to a higher gear, leading to in-memory computing (Fig. 2(b)) following the invention of memristive crossbar-arrays \citep{burrNeuromorphicComputingUsing2017, sebastianMemoryDevicesApplications2020}. Vector matrix multiplication (Fig. 2 (c)) is considered a key hardware booster in Deep Learning. The time and energy consuming task of vector matrix multiplication is performed in a memristive crossbar-array in which the input and output layer are interconnected by an array of weighted, checkboard arranged memristive devices  \citep{burrNeuromorphicComputingUsing2017, wangCrossbarBasedInMemoryComputing2020, liMemristiveCrossbarArrays2021}. Vector matrix multiplication is an example of how Deep Learning may benefit from the development of new electronic devices, e.g. memristors. Reservoir Computing (Fig. 2 (b)) was independently invented by Herbert Jäger and Wolfgang Maass and belongs to the general framework of Recurrent Neural Networks (RNN) \citep{jagerEchoStateApproach2001, maassRealTimeComputingStable2002}. In RNNs, a backpropagation through-time procedure is typically applied to adjust (train) the weights of the network to desired target functions. Here, a significant amount of time is required, with no certainty that the optimal weights will be set after learning. Thus in RC the reservoir consists of an ensemble of nonlinear elements coupled to one another. The reservoir projects incoming data and time series to a higher dimension that can be easily readout by conventional classifiers, in which the training is executed by means of a linear regression, for example. This reservoir can be either virtual or physical. These aforementioned reservoirs are designed like neural networks in which the connections are randomized but remain fixed during computation. Physical reservoirs are those which rely on natural systems exhibiting nonlinearity \citep{nakajimaPhysicalReservoirComputing2020, gauthierNextGenerationReservoir2021, milanoMateriaReservoirComputing2022}. 

The goal of analog computing is to mimic complex technical systems by means of electronic circuits which represent key system parameters as a set of voltage levels at nodes. Oscillatory computing (Fig. 2 (e)) refers to a subset of analog computing in which the oscillator frequencies and phases enrich the representation of information. Oscillatory systems are omnipresent in nature and engineering \citep{arenasSynchronizationComplexNetworks2008, pikovskijSynchronizationUniversalConcept2003, strogatzExploringComplexNetworks2001}. Technically, oscillators can be realized in numerous ways, such as in discrete or integrated semiconductor electronics, spin-torque devices, Josephson junctions, optical devices, or micro electro-mechanical systems(Schneider et al. 2018; Lequeux et al. 2016; Chen et al. 2020; Ignatov et al. 2016; X. Cheng et al. 2021; Feldmann et al. 2019a; C. Lenk, L. Seeber, and M. Ziegler 2020) \citep{schneiderUltralowPowerArtificial2018, lequeuxMagneticSynapseMultilevel2016, chenSpinOrbitTorque2020, ignatovSynchronizationTwoMemristively2016, x.chengCMOSIntegratedLowPower2021, feldmannAllopticalSpikingNeurosynaptic2019, lenk2020tuning}. In general, dynamical systems and their coupled oscillators may offer elegant solutions to compute HP-hard problems. Coupled oscillator networks have been successfully exploited in the field of pattern recognition \citep{kantnerDelayinducedPatternsTwodimensional2015, holzelPatternRecognitionSimple2011}. However, larger systems have not yet been successfully developed due to noise-stability problems and device constraints in the new class of compact oscillators based on VO$_2$ or NbO$_x$, for example \citep{shamsiHardwareImplementationDifferential2021, d.leeNbO2BasedFrequencyStorable2018}.

The term bio-inspired computing (Fig. 2 (f)) is only loosely defined. To a large extent, the computing primitives described above (see Fig. 1 and Fig. 2 (b)-(e)) are more or less biologically motivated. The Perceptron is a crude blue print of a neuron and is still today at the heart of Deep Learning systems \citep{rosenblattPerceptronProbabilisticModel1958}. In-memory computing is a strategy to abrogate the strict separation of the ALU and memory in digital computing, and is derived from biological information processing where logic and memory are blended. Neuromorphic processors contain circuits that can execute the Leaky Integrate-and-Fire dynamics of neurons, including the biologically motivated winner-take-all (WTA) principle, and introduces axon delays \citep{schumanSurveyNeuromorphicComputing2017, haslerFindingRoadmapAchieve2013}. Coupled oscillators imitate the orchestra of neural ensembles, i.e. the communication of separate brain regions which is considered to be the fundamental mechanism that explains perception \citep{singerVisualFeatureIntegration1995, varelaBrainwebPhaseSynchronization2001, buzsakiRhythmsBrain2006}. Cellular automata, for example, were introduced by John von Neumann to describe self-reproduction in biology \citep{mange2004self}. Probabilistic computing is based on Bayesian inference, which is closely related to the way humans make decisions \citep{alaghiSurveyStochasticComputing2013, parrComputationalNeuropsychologyBayesian2018}. Therefore, it is essential to declare precisely to what extent an artificially built system is bio-inspired and which biological pathway have been applied as design principles \citep{venkatesanBrainInspiredElectronics2022}.

\subsection{Novel electronic devices}

There is an ongoing effort to shrink silicon FETs to feature sizes below 5 nm. The FinFET structure has dominated CMOS technology since its invention in 1989 \citep{colingeFinFETsOtherMultigate2008}. Novel designs, such as GAAFET (Gate-All-Around) and MBCFET (Multi-Bridge-Channel), are serious candidates for next generation CMOS chips (see Fig. 3 (a)) \citep{radamsonStateArtFuture2020}. Aside from this ongoing improvement of conventional FETs, devices with novel functionalities and materials have been attracting considerable interest to implement novel computing architecture. Magnetic Josephson Junctions, photonic synapses, and bio-organic memories represent only a fraction of current development strategies \citep{schneiderUltralowPowerArtificial2018, feldmannAllopticalSpikingNeurosynaptic2019, robinModelingEmergentMemory2021, zhangOrganismicMaterialsNeumann2020}.  In Fig. 3 (b) to (d), unconventional nanoelectronics device structures are illustrated. In Fig. 3 (b), a memristive device structure is illustrated, comprising two electrodes separated by a memristive layer. In the same Figure, a qualitative I-V curve of a memristive device is shown alongside a sketch of a biological synapse (see also Fig. 4), highlighting that memristive devices are promising artificial synaptic counterparts due to their capability of presenting variable resistive weights in engineered neural networks \citep{bianStimuliResponsiveMemristive2021}. 

\begin{figure} [ht!]
	\centering
	\includegraphics[width=0.9\textwidth]{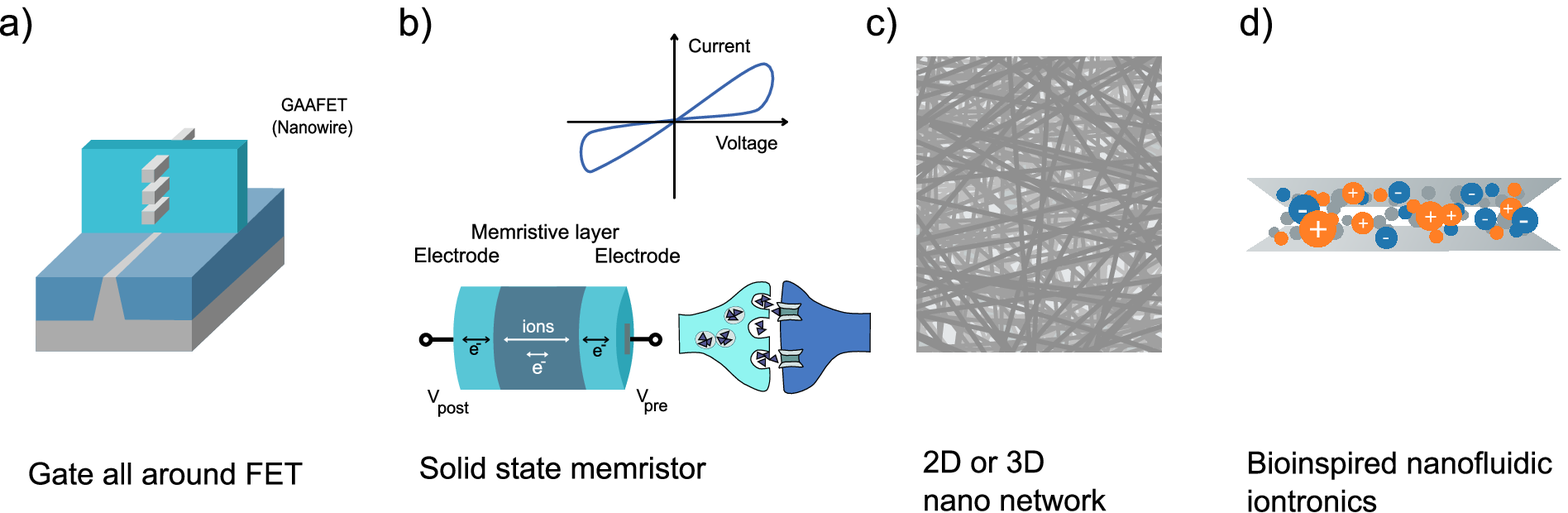}
	\caption{Schematics of four advanced device components. (a) 3D view graph of a GAAFET as applied in today’s latest digital processors \citep{radamsonStateArtFuture2020}, (b) sketch of a memristive device including a qualitative I-V curve and illustration of a synapse \citep{ielminiResistiveSwitchingFundamentals2016, xiaMemristiveCrossbarArrays2019, sunFutureMemristorsMaterials2021}, (c) cartoon of a 3D nanowire network \citep{stiegEmergentCriticalityComplex2012, pantoneMemristiveNanowiresExhibit2018, minnaiFacileFabricationComplex2017, mallinsonAvalanchesCriticalitySelforganized2019, loefflerTopologicalPropertiesNeuromorphic2020, zhuInformationDynamicsNeuromorphic2021, hochstetterAvalanchesEdgeofchaosLearning2021, kuncic2021neuromorphic}, (d) 3D cross-sectional graph of a fluidic memristive device adapted from \citep{robinModelingEmergentMemory2021} with permission.}
	\label{fig:advanced_device}
\end{figure}

One universal property of the memristive device concept is that the memristive state depends on previously induced charge flows, applied currents, or applied electric fields, thus storing a historically-determined resistance state. For details concerning resistive switching and the underlying physical-chemical mechanisms, we refer the reader to the references given in the figure caption (Fig. 3) and the overwhelming literature on the subject. \citep{ielminiResistiveSwitchingFundamentals2016, lanzaStandardsCharacterizationEndurance2021, bianStimuliResponsiveMemristive2021, wangRecentAdvancesVolatile2020}.

It is this concurrently complex and simple device concept, together with the tremendous predicted potential for breakthrough technologies in areas such as universal memories and novel non-Boolean computing schemes for cognitive electronic systems, that propels the research and development of memristors and memristor-based circuits worldwide. It is important to mention that, in contrast to the theoretically simplistic memristor concept, in practice the realization of memristive devices by modern thin film technology is a task littered with obstacles. Up until now, a huge number of experimental findings on memristor devices consisting of a broad variety of metal/insulator material combinations have been published, all of which show memristive I-V curves \citep{ielminiResistiveSwitchingFundamentals2016, wangRecentAdvancesVolatile2020, sunFutureMemristorsMaterials2021}. At first glance, it seems that the toolbox of resistive switching devices is ready for nearly any circuit application: simply pick a device concept and follow the extensive materials and methods laid out in the literature. However, a closer look at the fine details casts a dark shadow on this bright research field, leading to a harsh awakening based on hard facts. These “hard facts” are the requirements and boundary conditions set by the envisaged circuit applications, in which memristors must fit technologically, electronically, and economically. Currently, two main development avenues can be explored for memristive devices. The first focuses on resistive random access memories (RRAMs). It is believed that the zoo of today’s existing memory diversity can be replaced by a single (universal) memory concept. RRAMs are considered attractive candidates for universal memories because they: (i) show non-volatile data storage, (ii) can be densely integrated, (iii) are fast, and (iv) are cheap to produce. In particular, such a universal memory might attenuate the problem known as memory latency in modern digital computers \citep{iniewskiCMOSProcessorsMemories2010, banerjee2020}.
Besides the RRAM goal which may be categorized under the label “More Than Moore”, novel and very appealing computer architectures have been proposed in which memristors might play a vital role. Another main focus of possible memristive device applications may be associated with such catchphrases as: non-Boolean computing, bio-inspired information processing, neuromorphic engineering, or cognitive electronics \citep{zamarreno-ramosSpikeTimingDependentPlasticityMemristiveDevices2011, ranjanIntegratedCircuitMemristor2017, huangPatterningMetalNanowire2021, kasparRiseIntelligentMatter2021, wanEmergingArtificialSynaptic2019}. On the local, synaptic level, learning in nervous systems is explained by Hebb’s learning rule and Spike-timing dependent plasticity (STDP), amongst others \citep{biSynapticModificationsCultured1998}. STDP and other memory-related mechanisms observed in nervous systems, such as Long-term Potentiation (LTP) and Long-term Depression (LTD) \citep{blissLonglastingPotentiationSynaptic1973}, have been successfully mimicked by memristive devices \citep{ohnoShorttermPlasticityLongterm2011, winterfeldTechnologyElectricalCharacterization2018}. Moreover, traditional studies known from behaviorism, such as classical conditioning (e.g. Pavlov’s dog), anticipation, and optical illusions, were successfully realized experimentally by both single and pairs of memristive devices \citep{pershinExperimentalDemonstrationAssociative2010, zieglerElectronicVersionPavlov2012, bichlerPavlovDogAssociative2013, zieglerElectronicImplementationAmoeba2014, ignatovSynchronizationTwoMemristively2016}. The extent to which larger networks of memristive devices are able to mimic higher brain functions is still unknown. 

In Fig. 3 (c), a sketch of a nanowire network (NWN) is shown. NWNs have been successfully synthesized for various materials, such as metals, oxides, and semiconductors \citep{huangPatterningMetalNanowire2021, milanoMateriaReservoirComputing2022}. Nanowires show appealing features with respect to bio-inspired computing from the point of structure, topology, and inherent dynamics \citep{loefflerTopologicalPropertiesNeuromorphic2020, hochstetterAvalanchesEdgeofchaosLearning2021, zhuInformationDynamicsNeuromorphic2021, diaz-alvarezEmergentDynamicsNeuromorphic2019, pantoneMemristiveNanowiresExhibit2018}. In recent comprehensive reviews by Zhu et al. and Kuncic and Nakayama, hallmarks known from biological systems as small-world connectivity (topology) and self-organized criticality (dynamic) were addressed \citep{zhuInformationDynamicsNeuromorphic2021, kuncic2021neuromorphic}. Interestingly, brain-like avalanche effects have been observed in NWNs that exhibit dynamic features found in nervous systems \citep{pikeAtomicScaleDynamics2020, hochstetterAvalanchesEdgeofchaosLearning2021, beggsNeuronalAvalanchesNeocortical2003, chialvoEmergentComplexNeural2010}. Finally, we would like to emphasize that emergent neuromorphic materials and devices are not restricted to the solid state phase. In Fig. 3 (d), a sketch related to a nanofluidic device is shown. Bocquet and co-workers demonstrated by analysis and molecular dynamic simulations that ion transport across quasi–two-dimensional slits under an electric field displays memristive I-V curves, as well as spiking voltage patterns in accordance with the Hodkin-Huxley model of biological neurons \citep{robinModelingEmergentMemory2021, hodgkinQuantitativeDescriptionMembrane1952}. We would like to emphasizes that while these examples of NWNs and nanofluidics clearly demonstrate that the material “tool box” offers novel opportunities to implement higher brain function, its full potential has yet to be fully explored.

\section{Information Processing in Nervous Systems}

This perspective explores the role of information processing observed in nervous systems as a basis for the development of energy-efficient technological computing systems, and even the possibility of implementing higher brain functions in engineered systems. Nervous systems offer paradigms to improve energy-efficient artificial information processing units. The exploration of signal pathways in nervous systems shows us how evolution led to extremely energy-efficient signal processing units (nervous systems). For example, the human brain dissipates a power of only roughly 20 W to 25 W. This, in addition to the amazing capabilities of humans’ vision and hearing, reveals fascinating opportunities for autonomous vehicles or speech recognition. Hence, processing sensitive data in server clouds may lead to severe security concerns. The data of millions of cars in motion, including their controllability, falling into the wrong hands could lead to fatal attacks; Local data processing in an autonomous car with low power consumption is preferable.

Creatures are very well adapted to their specific ecological niche, a result of a hundred million years of ongoing evolution and the associated interaction between creatures and their environment throughout their life span \citep{martinezCircuitsBrainsOrigin2020, dobzhanskyNothingBiologyMakes1973, jacobEvolutionTinkering1977}. In particular, information pathways in nervous systems are prototypes for engineers to perform cognitive tasks in quasi-real time with extremely low power consumption \citep{poonNeuromorphicSiliconNeurons2011}. These features alone, and the information processing behind them, represent attractive models for entirely new computing architectures. In sections 4 A. and 4 B., local and global aspects of information processing mechanisms are presented, respectively. Distinct differences between digital computing and information pathways in biological systems are highlighted in the framework of topology and dynamics to motivate the concept of artificial spatio-temporal networks, as subset of the field of bio-inspired information processing. In sections 4 C. (Phylogenies and Ontogenesis) and 4 D. (Homeostasis), we underline important hallmarks of information processing in biological systems which have so far only been partly considered for artificial systems. Note that in chapter 4, we do not address how such mechanisms can be established in electronics: This is the subject of chapter 5, where several approaches are proposed to implement an artificial spatio-temporal network. It is not our goal to develop another pattern recognition system but to address the fundamental question: “To what extent can higher brain functions be reproduced in artificial systems?”. We believe that essential information pathways in biology have been to a large extent overlooked, as detailed in this perspective. One important difference between artificial spatio-temporal networks and contemporary AI and neuromorphic engineering is that essential growth mechanisms observed in nervous systems are exploited as a guideline in the former.

\subsection{Local Aspects of Information Processing in Nervous Systems}

In contrast to current clock-driven Boolean Turing machines, information processing in biological nervous systems is characterized by highly parallel, energy-efficient, and adaptive architecture \citep{backusCanProgrammingBe1978, turingCOMPUTINGMACHINERYINTELLIGENCE1950, rueckertBrainInspiredArchitecturesBrainInspiredArchitectures2016}. When it comes to pattern recognition, failure tolerance, and cognitive tasks, even simple creatures outperform supercomputers, in particular regarding power dissipation. Fundamental building blocks leading to such remarkable properties exploit neurons as central processing units, which are interconnected by synapses to form a complex dynamical three dimensional network, the connectome \citep{seungConnectomeHowBrain2012}. 
In Fig. 4, the structure of a neuron is sketched, including the soma, dendrites, the axon, and connections to other neurons by synapses.

\begin{figure} [ht!]
	\centering
	\includegraphics[width=0.7\textwidth]{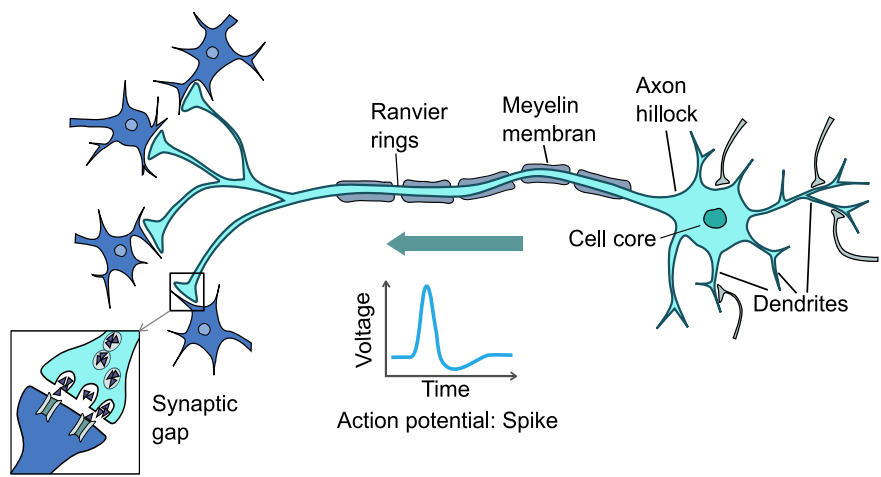}
	\caption{Blueprint of a neuron including an enlarged sketch of a synapse and the illustration of a single action potential, a spike.}
	\label{fig:neuron}
\end{figure}

An action potential (spike) is defined as a sudden transitory and propagating change in the resting potential across a membrane. Action potentials sent from presynaptic neurons are received via the dendrites and synapses of the postsynaptic neurons. Those signals are integrated within the cell body of the postsynaptic neuron. When a threshold potential is reached, the neuron generates a new spike or a sequence of new spikes at the axon hillock that are transmitted via the axon to a postsynaptic neuron. This entire process is called Leaky Integrate-and-Fire (LIF). The term leaky reflects the fact that the cell membrane is not a perfect electrical insulator. Numerous LIF models, such as the FitzHugh–Nagumo, Morris–Lecar, or Hindmarsh–Rose models, have been developed to address different aspects of the biological substrate \citep{dayanTheoreticalNeurosciencesComputational2001, gerstnerSpikingNeuronModels2002, izhikevichSimpleModelSpiking2003, izhikevich2010hybrid}.  Depending on the electrical activity of two connected neurons, the connection strengths (the weights) can become weaker or stronger. This is at the heart of Donald E. Hebb’s learning rule, who first recognized that “Neurons which fire together wire together” \citep{hebbOrganizationBehavior2005}. On the biochemical level the variable strength is explained by the amount of neuro transmitters (vesicles) which are released into the synaptic cleft. 

From an engineering point of view, nervous systems process information in such a way that silicon technology, the holy grail of modern digital computing strategies, seems to be outmatched. For example, electronic components and circuits, such as transistors, memories and processors, are optimized for small parameter spreads to run at GHz clock frequencies under a precise pulse timing \citep{iniewskiCMOSProcessorsMemories2010, j.m.veendrickNanometerCMOSICs2017}. In particular, they exploit nanosecond signal pulses that travel at nearly the speed of light along well-ordered transmission lines that connect different system parts in an essentially two-dimensional topology. In contrast, information pathways in nervous systems are characterized by highly irregular tissue consisting of neurons, synapses, and axons. Low conduction velocities on the order of several m/s lead to pronounced signal retardation, i.e. delays. In Fig. 5, characteristic timescales of CMOS processors and nervous systems are compared. In digital computing, the pulse duration is below a ns, and the signal transmission velocity is at nearly the speed of light. In nervous systems, the corresponding values are 3.5 ms for the pulse duration of an action potential or spike, and a few tenths of a ms for the transmission of a spike along myelinated axons \citep{kandelPrinciplesNeuralScience2013}. Whereas the clock frequency of a modern Si processor is about 5 GHz, human EEG brain waves range from below 1 Hz to a few 100 Hz \citep{heScalefreeBrainActivity2014, buzsakiNeuronalOscillationsCortical2004, buzsakiRhythmsBrain2006}. This represents a six orders of magnitude discrepancy between technical and biological parameters. These facts alone point towards fundamental differences between information processing in digital computing and those in natural nervous systems.

\begin{figure} [ht!]
	\centering
	\includegraphics[width=0.5\textwidth]{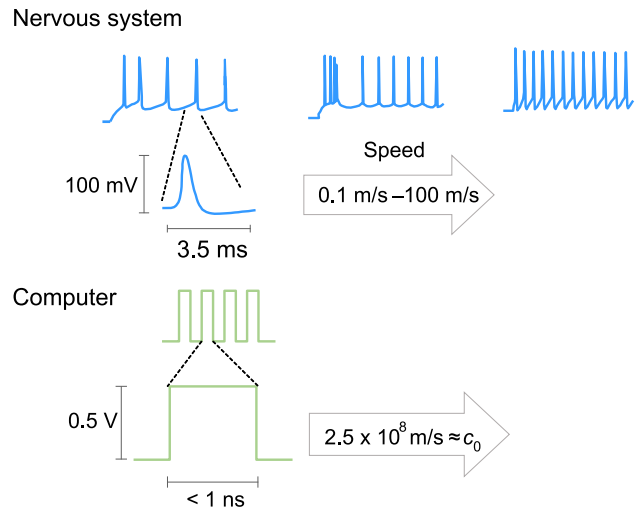}
	\caption{Comparison between pulse transmission speed, pulse duration, and voltage amplitude in nervous systems and digital computing. The sequence of action potentials were adapted with permission from Fig. 1 of ref. \citep{izhikevichSimpleModelSpiking2003}}
	\label{fig:pulse_transmission}
\end{figure}

\subsection{Global Aspects of Information Processing in Nervous Systems}

Nervous systems are considered to be time-varying networks in which spike-dynamics and cellular morphology are intricately linked and reciprocally interwoven \citep{fornitoFundamentalsBrainNetwork2016, martinezCircuitsBrainsOrigin2020, winfreeGeometryBiologicalTime2001, thompsonGrowthForm1992}. 

Information processing in nervous systems applies a broad range of structurally and temporally related phenomena \citep{kandelPrinciplesNeuralScience2013, nassimLessonsLobsterEve2018, sterlingPrinciplesNeuralDesign2015}. At the level of individual synapses, neurons, and axons, the formation and transmission of action potentials (“spikes”) are reasonably well understood. However, a look at the mesoscopic and macroscopic level of the three-dimensional neuronal network leads to an entirely different assessment. Although groundbreaking progress has been reported on \textit{in vivo} and \textit{in vitro} techniques over the last decades, the nervous system’s spatiotemporal information processing is still not well understood \citep{kleinfeldControlledOutgrowthDissociated1988, kriegeskorteCognitiveComputationalNeuroscience2018, schroeterEmergenceRichClubTopology2015, vandenheuvelExploringBrainNetwork2010}. The biochemical mechanisms that explain higher brain functions at the cellular level, such as awareness, perception, and in particular consciousness, remain elusive \citep{mackeyOscillationChaosPhysiological1977, engelTemporalBindingBinocular1999, dehaeneCognitiveNeuroscienceConsciousness2001, bullmoreComplexBrainNetworks2009, bassettUnderstandingComplexityHuman2011, melloniMakingHardProblem2021}. Nonetheless, neuroscientists were able to identify basal mechanisms that define the fundamental platform of the unique and marvelous nervous system’s information processing. Characteristic features, such as STDP \citep{biSynapticModificationsCultured1998, gerstnerSpikingNeuronModels2002, markramSpikeTimingDependentPlasticityComprehensive2012}, stochastic firing and bursting of neurons in the hundred Hz range, recurrent network structures, and aspects of oscillatory synchrony in larger neuronal ensembles \citep{ernstSynchronizationInducedTemporal1995, hoppensteadtWeaklyConnectedNeural1997, buzsakiRhythmsBrain2006, arenasSynchronizationComplexNetworks2008, galiziaNeurosciencesMoleculeBehavior2013, gerstnerNeuronalDynamicsSingle2014, amilOrganizationIdentificationSolutions2015, strogatzNonlinearDynamicsChaos2015, wattsCollectiveDynamicsSmallworld1998, uhlhaasNeuralSynchronyCortical2009} are essential ingredients in biologically-based information processing. Moreover, factors related to the close interaction of a nervous system with its environment, i.e. external stimuli, are of crucial importance \citep{beerDynamicalApproachesCognitive2000}. Therefore, neuronal design principles provide a model for bio-inspired computing systems, which are diametric to development strategies in present binary IT, including GHz clock frequencies, near-light-speed signal transmission, and clearly separated from logic and memory \citep{hofflingerNewVistasNanoelectronics2016, j.m.veendrickNanometerCMOSICs2017}. 

Beyond that, we would like to emphasize that information-related aspects of nervous systems during evolution (phylogenies), along with their individual development throughout their lifetime (ontogenesis), provide a promising model from which novel electronic architectures may be designed. In the animal kingdom, the intricacy of nervous systems varies tremendously between single- and multi-cellular organisms, and the human brain with its billions of interconnected neurons \citep{kandelCellularBasisBehavior1976, nakagakiMazesolvingAmoeboidOrganism2000, azevedoEqualNumbersNeuronal2009, boschHydraPolypNothing2010, bieleckiSwimPacemakerResponse2012, naumannReptilianBrain2015, boschBackBasicsCnidarians2017, dupreNonoverlappingNeuralNetworks2017, giezNeuronsInteractMicrobiome2021}. For the sake of completeness, we would like to specify that the existence of cognitive functionalities in entities without a nervous system, such as plants or the acellular slime mold Physarum polycephalum, is currently heavily debated. For interested readers, more detailed information can be found in the following references \citep{vallverduSlimeMouldFundamental2018, gaglianoExperienceTeachesPlants2014, adamatzkyAdvancesUnconventionalComputing2017, stepney2017inspired}. 

Despite their different cognitive capabilities, neurons and nervous systems present many common features in all creature, such as synapses, signal transmission lines (axons), and action potentials (spikes), that act as basic information building blocks. While the term morphology defines the real structure of a nerve net, the topology of a net is more abstract, related to important theoretical graphical parameters that define the connectome of a nervous system \citep{barabasiEmergenceScalingRandom1999, barabasiNetworkScience2013, bullmoreComplexBrainNetworks2009, fornitoFundamentalsBrainNetwork2016}. The connectome is considered to be the canonical state describing the cellular wiring diagram of a nerve net. Edges, nodes, cluster coefficients, characteristic path lengths, hubs, and motifs determine the topological quality of a net, for example. An unraveling of the micro- and macro-connectome and nervous system dynamics offer a suitable model for the next generation of bioinspired hardware electronics \citep{sterlingPrinciplesNeuralDesign2015}. 

\begin{figure} [ht!]
	\centering
	\includegraphics[width=0.9\textwidth]{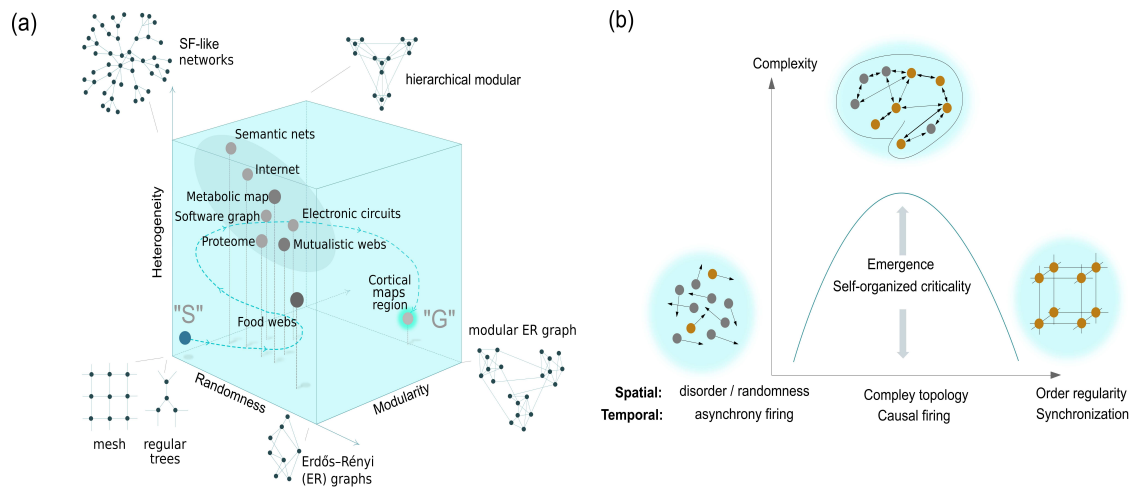}
	\caption{Network cube and complexity: (a) classification of various networks. The dashed blue line illustrates a fictive, guided “walk” through the cube, starting from “S” and ending at the goal “G”. In this way, it will be possible to push the network properties of neuromorphic circuits towards those of cortical maps (see lower right-hand corner in the cube). This approach is part of an artificial spatio-temporal networks. (b) Qualitative illustration of the complexity term. From a spatial (topological) and temporal (dynamical) point-of-view, a complex system is neither completely random nor entirely ordered, but exhibits a state in between. Figures 6 (a) and (b) are adapted with permission from  Solé and Valverde 2004 \citep{soleInformationTheoryComplex2004} and Huberman and Hogg 1986 \citep{hubermanComplexityAdaptation1986}, respectively.}
	\label{fig:networkcube}
\end{figure}

The network cube (Fig. 6 (a)) classifies a number of different nets according to theoretical attributes, including randomness, modularity, and heterogeneity  \citep{soleInformationTheoryComplex2004, spornsNetworksBrain2011}. Interestingly, in this framework, cortical maps (lower right corner of the cube) extracted from the structural properties of nervous systems are somewhat isolated from all other nets, which are located in the upper left corner of the cube. In Fig. 6 (b), dynamical complexity (y-axis) is described as a state between complete asynchrony, with independent, random firing of the individual oscillators, and complete synchrony, with all oscillators firing in phase \citep{hubermanComplexityAdaptation1986}. Between these two extremes, system dynamics can be characterized by a complex and time-varying interaction of the oscillatory ensemble. This regime exhibits features of self-organized criticality (SOC) typically observed at and near phase transitions and might be identified by avalanches of firing neuron ensembles \citep{bakSelforganizedCriticality1988, beggsNeuronalAvalanchesNeocortical2003, hesseSelforganizedCriticalityFundamental2014, shewAdaptationSensoryInput2015, mallinsonAvalanchesCriticalitySelforganized2019, cramerControlCriticalityComputation2020}. Avalanche behavior is common in many physical phenomena, such as magnetic systems, earthquakes, and brain dynamics at the critical region of phase transitions, and were first described by Bak et al. \citep{bakSelforganizedCriticality1988}. The common feature of all these systems is slow external driving, causing an intermittent, widely distributed response. Avalanches appear in very different sizes, often distributed in the form of power laws. As known from statistical physics, power laws imply the absence of a characteristic scale, a property observed close to a critical point. When describing the dynamics in a nervous system using the SOC and brain-like avalanches models, the type of phase transition associated to each term must be clearly defined. For example, SOC and brain-like avalanches in NWNs (see section 3 B. and Fig. 3 (c)) are related to non-activity – activity phase transitions. In the context of firing neuron ensembles in a brain, SOC and avalanches may describe a temporal phase transition between the asynchronous and synchronous states \citep{hesseSelforganizedCriticalityFundamental2014}. In other words, a system could be in the supercritical state (above the critical point) in an inactive -active phase transition, while remaining subcritical (below the critical point) with respect to the asynchronous-synchronous phase transition. However, such phase transitions are not necessarily exclusive and might appear simultaneously in the brain, or the mechanisms could even be interwoven.
So far, while nanoparticle networks and NWNs have been studied in context with their activity pattern, neuron-like oscillatory components have yet to be considered. The orchestra of firing neuron ensembles is considered a key underlining mechanism in understanding the binding problem, i.e. the capability of the brain to integrate (bind) different sensory inputs. For example, such a process can occur in the visual system when forming a unified perception of the environment \citep{vondermalsburgWhatWhyBinding1999, singerVisualFeatureIntegration1995, uhlhaasNeuralSynchronyCortical2009, engelDynamicPredictionsOscillations2001, schechterHowBrainGets1996, sheffieldBindingSolution2015, engel1991interhemispheric}. Suggestions on ways to include relaxation-type oscillators to mimic the LIF features of biological neurons and the state of SOC are presented in chapter 5. 
Finally, the topological and temporal dynamics of the regime are extremely sensitive to external distortions (stimuli) at the critical point, allowing the system to respond in numerous ways to external stimuli \citep{chialvoEmergentComplexNeural2010, chialvoAreOurSenses2006, kinouchiOptimalDynamicalRange2006, beggsNeuronalAvalanchesNeocortical2003, shewInformationCapacityTransmission2011}. In biological terms, this means that the system can easily adapt to risky environmentally-driven situations. The manifold brain states available near the point of criticality offer a wide repertoire of means to react in a reasonable way to external tasks imposed by the environment. In extreme situations, this improves the chances of survival and is of evolutionary importance.

\subsection{Phylogenies and Ontogenesis}

The origin of bio-inspired computing can be best drawn from the two following neuroscience quotations:

(1)
Gilles Laurent pointed out the common evolutionary heritage of living organisms. His contribution “Shall We Even Understand the Fly’s Brain? (see: 23 Problems in Systems Neuroscience edited by J. L. van Hemmen and T. J. Sejnowski, Chapter 1, page 3, \citep{hemmen23ProblemsSystems2006}) states: “When it comes to computation, integrative principles, or “cognitive” issues such as perception, however, most neuroscientists act as if King Cortex appeared one bright morning out of nowhere, leaving in the mud a zoo of robotic critters, prisoners of their flawed designs and obviously incapable of perception, feeling, pain, sleep, or emotions, to name but a few of their deficiencies.” \citep{laurentShallWeEven2006}.

(2)
Martijn P. van den Heuvel et al., made in “The Neonatal Connectome During Preterm Brain Development” the following statement: “The adult cerebral brain network is the result of a complex developmental trajectory. From the prenatal formation of the first neurons, throughout the first years of life and all the way into late adolescents, the brain undergoes an elaborate developmental trajectory.” \citep{vandenheuvelNeonatalConnectomePreterm2015}.

How are these sayings so important for the design of novel bio-inspired computing primitives?  The general idea behind these two quotations is the concept of development. Quotation (1) by Gilles Laurent highlights evolutionary development, the phylogenies of species, and their relevance to the emergence of the human cortex. This bottom-up approach favors the study of less complex creatures that appeared early during evolution, laying the foundation for much more complex nervous systems in vertebrates \citep{naumannReptilianBrain2015}. In particular, information processing strategies throughout evolution and in completely different species are astonishingly similar, if not the exact same. For example, the basic ingredients of information processing (neurons, synapses, and action potentials, as described in chapter 3)  in the nervous systems of squids and macaques are hardly distinguishable from one another. Although François Jacob addressed the random and playful character of evolution by the phrase “Nature is a tinkerer, not an inventor” \citep{jacobEvolutionTinkering1977}, evolution can be somewhat conservative in the sense that similar structural and dynamical features appear in very different species throughout phylogenies. This justifies the investigation of information pathways in simpler, easier-to-understand organisms in order to comprehend higher brain functions in more complex vertebrates. A famous example is the research of Eric Kandel on the snail Aplysia, relating physiological signaling with behavior \citep{kandelCellularBasisBehavior1976}. Studying the neural design of biological species with only a few hundred or thousands neurons is a fruitful ansatz to develop novel computing primitives \citep{brennerGeneticsCAENORHABDITISELEGANS1974, kaiserEvolutionDevelopmentBrain2011, sterlingPrinciplesNeuralDesign2015, boschBackBasicsCnidarians2017, dupreNonoverlappingNeuralNetworks2017, lovasEnsembleSynchronizationReassembly2021, giezNeuronsInteractMicrobiome2021}.  We will come back to this issue in chapter 5.

While phylogenetics addresses the development and evolution of groups of similar species, ontogenesis is the study of how an individual member of a species develops as it ages.  In quotation (2), Martijn P. van den Heuvel and coworkers underline the intriguing mechanisms of nervous systems development in humans, from conception to late adolescence. We propose that ontogenesis and their functional ingredients could serve as an essential guideline for novel computing primitives. To support this argument, we describe here the fundamentals of ontogenesis in the human nervous system, including the importance of external stimuli during development. Physiology, neurobiology, and behavioral science provide overwhelming experimental evidence showing that the conditions during the growth and regeneration of neuronal nervous systems under external stimuli are of central importance \citep{heldMovementproducedStimulationDevelopment1963, huttenlocherMorphometricStudyHuman1990, dehaene-lambertzInfancyHumanBrain2015, kaiserChangingConnectomesEvolution2020, gauthierNextGenerationReservoir2021, wieselSINGLECELLRESPONSESSTRIATE1963, paredesExtensiveMigrationYoung2016}. Both the formation and elimination of nerve cells, synapses, and axonal connections occur frequently during the first stages of brain development, belonging to a very creative process that shapes the nervous system to be well-adapted for future environment-related tasks. In addition to the creation of neurons and axons, pruning (programmed cell death or apoptosis) and axonal rewiring are both essential and expedient mechanisms. Finally, myelination of axons is an essential step to improve the nervous system’s performance by shaping and optimizing the signal transfer time between neurons and distributed brain areas. Gerald Edelmann coined the expression “Neuronal Darwinism” to highlight the striking parallels between evolution and brain development(Edelman 1987; Tononi, Sporns, and Edelman 1994; Edelman and Tononi 2001; Van Ooyen and Butz-Ostendorf 2017) \citep{edelmanNeuralDarwinismTheory1987, tononiMeasureBrainComplexity1994, edelmanUniverseConsciousnessHow2001, vanooyenRewiringBrainComputational2017}. Neurons, synapses, and axonal connections grow lavishly at first, a growth that is controlled by the genome, epi-genome, and stochastic factors. Subsequent structural shaping and elimination, often called blooming and pruning, are largely influenced by the interaction of the entire nervous system with environmental stimuli, and the nervous system’s subsequent reaction \citep{beerDynamicalApproachesCognitive2000, huttenlocherNeuralPlasticityEffects2002, henschCriticalPeriodPlasticity2005}. 
There have been attempts in the past to design materials and systems that mimic biological information processing, dubbed “evolvable hardware” and “evolution-in-materio” \citep{broersmaNascenceProjectNanoscale2012, stepney2017inspired, stiegEmergentCriticalityComplex2012, millerEvolutioninmaterioEvolvingComputation2014, adamatzkyAdvancesUnconventionalComputing2017}. This work has been recently extended to novel, transistor-based devices by Baek, et al. \citep{baekIntrinsicPlasticitySilicon2020}. Although the findings are very promising, basal spatiotemporal and topologically-relevant mechanisms have not been reproducible in electronics hardware so far. In both biological and artificial systems, the connection between these mechanisms should be worked out with regard to the required complexity and functionality (see Fig. 6). 

Neural network growth in nervous systems has been studied in-depth both theoretically and experimentally \citep{engertDendriticSpineChanges1999, kaiserSpatialGrowthRealworld2004, kaiserHierarchyDynamicsNeural2010, kouiderNeuralMarkerPerceptual2013, huttmarc-thorstenPerspectiveNetworkguidedPattern2014, vandenheuvelNeonatalConnectomePreterm2015, kaiserChangingConnectomesEvolution2020, hiesingerBrainWiringComposite2021, hiesingerSelfassemblingBrainHow2021}. In particular, the early stages of nervous system growth under external stimuli is of critical importance for the healthy development of mature creatures \citep{wieselSINGLECELLRESPONSESSTRIATE1963, ardielImportanceTouchDevelopment2010, seungConnectomeHowBrain2012}. It is known that both external stimuli and genetic factors have tremendous impact on the emergence of functional neural circuits that determine behavior during critical periods of cortical region growth. \citep{huttenlocherNeuralPlasticityEffects2002, henschCriticalPeriodPlasticity2005, lohmannDevelopmentalStagesSynaptic2014, hiesingerSelfassemblingBrainHow2021}.
Cell overproduction and subsequent attrition are likewise important for nervous system development \citep{huttenlocherNeuralPlasticityEffects2002, sanesDevelopmentNervousSystem2006, rumpelDynamischeKonnektom2016}. Morphological aspects, connectivity, growth, regeneration, and the impact of neuronal activity-related spike-based synchronization mechanisms in neuronal network ensembles serve as models for novel electronics \citep{fauthChapter16LongTerm2017, arenasSynchronizationComplexNetworks2008, uhlhaasNeuralSynchronyCortical2009, buzsakiRhythmsBrain2006, singerConsciousnessStructureNeuronal1998, buzsakiNeuronalOscillationsCortical2004, bassettUnderstandingComplexityHuman2011}. Clear evidence of structural dendritic spine plasticity is shown in a series of photographs taken over a few days in Fig. 7, demonstrating that spines grow and shrink depending on external, touch-related stimuli in mice \citep{holtmaatExperiencedependentCelltypespecificSpine2006}.

\begin{figure} [ht!]
	\centering
	\includegraphics[width=0.7\textwidth]{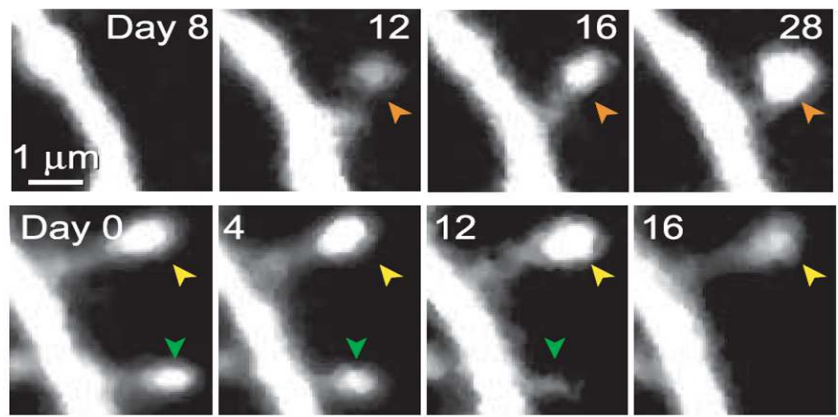}
	\caption{Structural plasticity. \textit{In vivo} time snapshots of the appearance and disappearance of dendritic spines in the mice barrel cortex. The top row shows the growth of a persistent spine between days 12-28 (orange arrows). The bottom row shows examples of spine retraction (yellow arrows between days 0-16, and green arrows between days 0-12). Figure from \citep{holtmaatExperiencedependentCelltypespecificSpine2006} with permission. These structural changes were correlated to external stimuli applied by whisker trimming in mice.}
	\label{fig:plasticity}
\end{figure}

A look at a few growth parameters underlines the importance of understanding biological networks during development. A two-year-old human toddler exhibits the maximum number of neurons and synapses of our species, roughly a factor of two more than a fully grown adult. If we estimate 170 billion neurons \citep{huttenlocherNeuralPlasticityEffects2002, azevedoEqualNumbersNeuronal2009} with 10$^3$ synapses per neuron, a two-year-old human carries 170 x 10$^{12}$ synapses. We assume the total axon length of a toddler to be about 850,000 km (https://aiimpacts.org/transmitting-fibers-in-the-brain-total-length-and-distribution-of-lengths/). The time between egg fertilization to the age of two is 1000 d or 8,64 x10$^7$ s. This leads to an average net growth of roughly 2000 neuron/s, 2 million synaptic interconnections/s and an axon growth rate of about 10 m/s! These measures alone unambiguously demonstrate the overwhelming significance of network growth in humans, particular during childhood \citep{kolbDevelopmentChildBrain2009, paredesExtensiveMigrationYoung2016}. Moreover, we believe that such a tremendous development is an interesting template for novel computing architectures. It might be an essential building block to achieve higher brain functionalities in artificial systems, and constitutes a key aspect artificial spatio-temporal networks.

Fig. 8 shows several snapshots taken during human development, where the excessive growth of neurons between the ages of one month to two years is clearly visible. Interestingly, between the ages of two to four years, neuron pruning leads to reduced neuron density. While the net neuron density during adulthood is rather constant, blooming and pruning still continue to occur, albeit at a much lower rate \citep{vanooyenRewiringBrainComputational2017}.

\begin{figure} [ht!]
	\centering
	\includegraphics[width=0.7\textwidth]{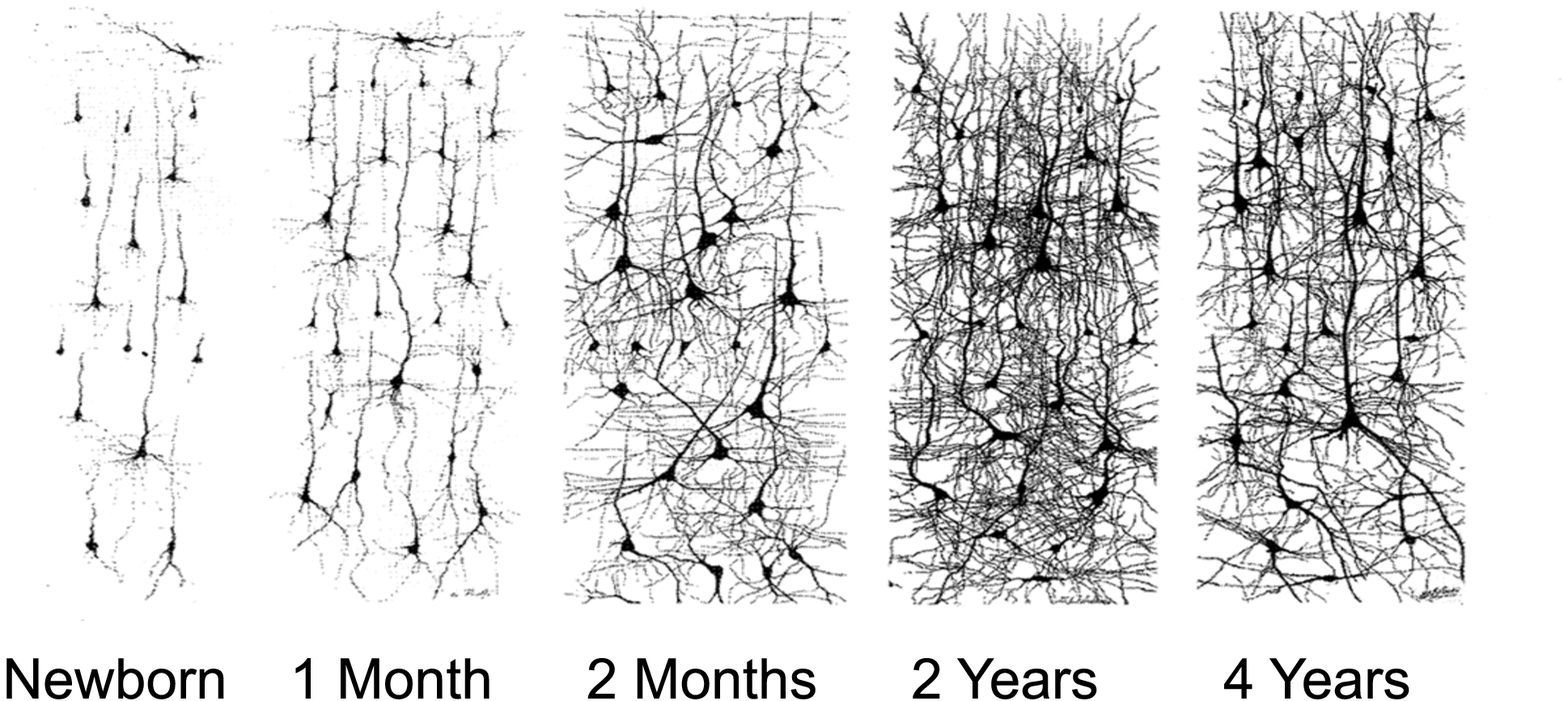}
	\caption{Blooming and pruning of nerve cells in young humans \citep{seungConnectomeHowBrain2012} and J. Conel, “The Post-Natal Development of the Human Cerebral Cortex,” Harvard University Press, Cambridge, 1939-1967. Adapted from \citep{seungConnectomeHowBrain2012}}
	\label{fig:blooming_pruning}
\end{figure}

From the postnatal phase up to the age of around two years, our central nervous system is characterized by enormous development and permanent remodeling, while being simultaneously subject to an exuberant amount of external stimuli via our senses \citep{beerDynamicalApproachesCognitive2000, kouiderNeuralMarkerPerceptual2013, henschCriticalPeriodPlasticity2005, vandenheuvelNeonatalConnectomePreterm2015}.  Genetics, stochastics, and external stimuli (in other words nature and nurture) define who we are and strongly influence higher brain function during adulthood, including perception, awareness, and consciousness. 

In Fig. 9, windows of plasticity in human brain development are sketched \citep{henschCriticalPeriodPlasticity2005, henschReopeningWindowsManipulating2012}. Even in much simpler creatures (e.g. the worm C.-elegans), external stimuli play an essential role in the healthy development of the nervous system \citep{ardielImportanceTouchDevelopment2010}.

\begin{figure} [ht!]
	\centering
	\includegraphics[width=0.7\textwidth]{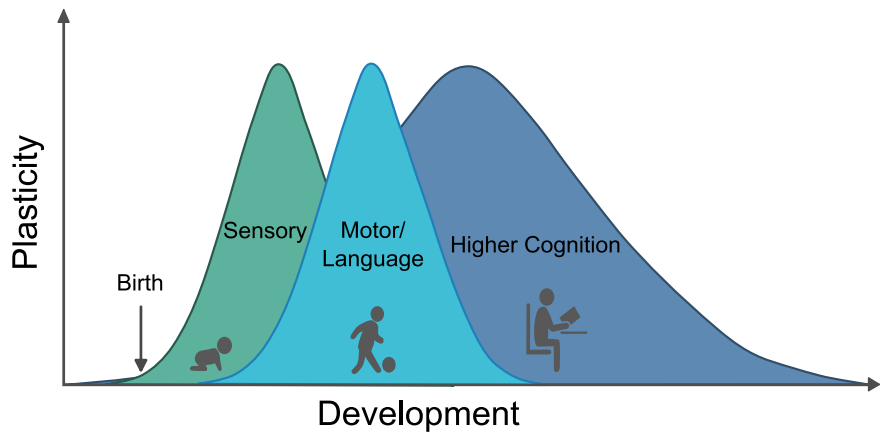}
	\caption{Illustration of critical or sensitive periods during the first years after birth for humans. The three periods present (from left to right) the development of sensory pathways, motor skills, and higher cognitive functions. Adapted from \citep{henschCriticalPeriodPlasticity2005, henschReopeningWindowsManipulating2012} with permission.}
	\label{fig:sensitive_periods}
\end{figure}

These windows for sensing, motor skills, and higher cognition are also called critical periods. They reflect the tremendous rearrangement of the human brain during early childhood, accompanied with enormous learning capabilities. It is interesting to assign the above estimated growth parameters and the appearance of critical periods to human altriciality.  Altriciality refers to the way creatures are born completely incapable of caring for themselves (Dunsworth et al. 2012). Hence, at the moment of birth (eye opening), a sudden rush of external stimuli, in particular vision, meets a premature nervous system still under heavy construction, reconstruction, and growth, in the case of humans. The concomitant occurrence of environmental input, nervous system growth, and close interaction between the nervous system and its environment may explain the huge plasticity and learning capabilities during these first years. This development seems to be essential to form higher brain functions \citep{dehaene-lambertzInfancyHumanBrain2015, vandenheuvelNeonatalConnectomePreterm2015}. Although it might be incredibly difficult to mimic such basal neurobiological mechanisms in engineered systems, nervous system development and growth cannot be neglected in establishing higher brain functions in artificial systems. Attempts to achieve this goal are proposed in chapter 5. 

\subsection{Homeostasis}

As in section 4 C., we begin with the following sequentially-labelled neuroscience quotation:
(3) Arjen van Ooyen and Markus Butz-Ostendorf emphasized the role of homeostasis on p.133 of their contribution (see: The Functional Role of Critical Dynamics edited by Nergis Tomen, J. Michael Herrmann, and Udo Ernst \citep{tomenFunctionalRoleCritical2019}: “In conclusion, during development, homeostatic structural plasticity can guide the formation of synaptic connections to create a critical network that has optimal functional properties for information processing in adulthood.” \citep{ooyen2019homeostatic}.

Roughly speaking, is homeostasis a kind of counteracting mechanism to network plasticity, and thus an important factor to ensure network robustness and stability? As will be discussed below in more detail, homeostasis comprises dynamical and morphological components, and is thought to explain how a nervous system stabilizes (itself) near the point of criticality \citep{bruttRoleExcitationInhibition2021}. In other words, homeostasis addresses the term “self” in SOC. The role of homeostasis as a stabilizing factor in neural networks is amply described in a huge number of publications, with only a few mentioned here (Abbott 2003; Turrigiano 2012; C. Tetzlaff et al. 2010; Stepp, Plenz, and Srinivasa 2015; Fauth, Wörgötter, and Tetzlaff 2017; van Ooyen 2017; Ma et al. 2019) \citep{abbottBalancingHomeostasisLearning2003, turrigianoHomeostaticSynapticPlasticity2012, tetzlaffSelfOrganizedCriticalityDeveloping2010, steppSynapticPlasticityEnables2015, fauthChapter16LongTerm2017, vanooyenRewiringBrainComputational2017, maCorticalCircuitDynamics2019}.  In homeostatic structural plasticity, all incoming synapses of a cell are modified to stabilize the neuronal activity around a particular level (set point), and reflect a negative feedback loop between neuronal activity and connectivity \citep{vanooyenRewiringBrainComputational2017, kehayas2017structural, tien2018homeostatic, turrigianoHomeostaticSynapticPlasticity2012}. The fundamental principle of homeostasis is sketched in Fig. 10.

\begin{figure} [ht!]
	\centering
	\includegraphics[width=0.48\textwidth]{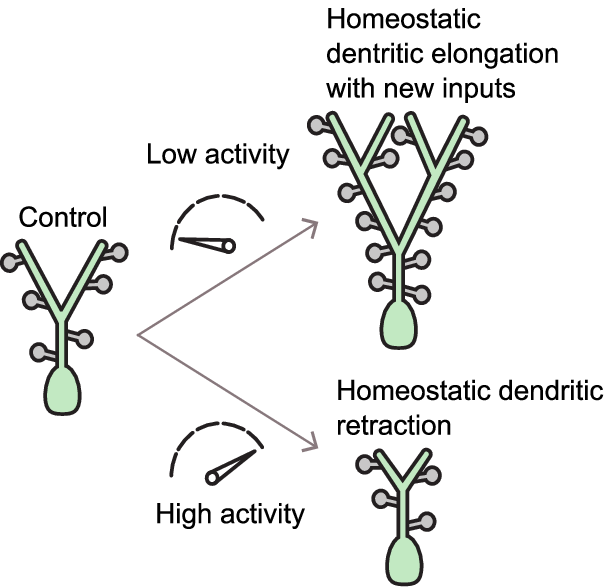}
	\caption{Illustration of homeostasis in a nervous system at the neuron level. Adapted from  \citep{butz-ostendorfChapterLesionInducedSynaptic2017} and \citep{tien2018homeostatic} with permission.}
	\label{fig:homeostasis}
\end{figure}

Higher firing (dynamic component) of a neuron results in spine deletion (morphological component), whereas reduced firing supports spine formation, keeping the average electrical activity at a set-point, potentially stabilizing the global activity of the entire neural ensemble near the desired critical state, i.e. the state with the largest dynamic range for information processing \citep{chialvoAreOurSenses2006, kinouchiOptimalDynamicalRange2006, shewInformationCapacityTransmission2011, butz-ostendorfChapterLesionInducedSynaptic2017}. While this model appears attractive at first glance, it raises a fundamental question in neuroscience: “How can an individual, local neuron in a huge nervous system access the global network state in order to orientate its own activity accordingly?” \citep{hesseSelforganizedCriticalityFundamental2014, fornitoFundamentalsBrainNetwork2016}, or in other words, what defines the activity set-point? This is an example of the poorly understood relation between local, mesoscopic, and global mechanisms in nervous systems. 

In chapters 3 and 4, we presented various basal local and global information pathways in nervous systems. In the following chapter, we will suggest a number of strategies with the goal of implementing higher brain functions in artificial systems \citep{millerEvolutioninmaterioEvolvingComputation2014}.

\section{Artificial Spatio-temporal Networks}

At this point, an obvious and understandable question might be: Is the goal to achieve higher brain functions in artificial systems possible at all or, more precisely, to what extent can the intriguing and complex biological mechanisms described in Chapters 3 and 4 be merged into a novel computing primitive? How close is neuroscience to understanding higher brain functions and to what extent can the plethora of phenomena set by materials and engineering designs strategies enable mental functions in artificial systems?  

Here we discuss possible ways and limitations of using artificial systems to mimic biological fundamentals, including topological and dynamical aspects, such as phylogenies, ontogenies, homeostasis, SOC, memory, oscillatory orchestra (synchrony), and so on. Nonetheless, we are aware that fundamental limits which may impede consciousness in engineered systems. It would be interesting, however, to identify and define those limits.

In Fig. 11, considerations set by materials science and design strategies are illustrated.

\begin{figure} [h]
	\centering
	\includegraphics[width=0.9\textwidth]{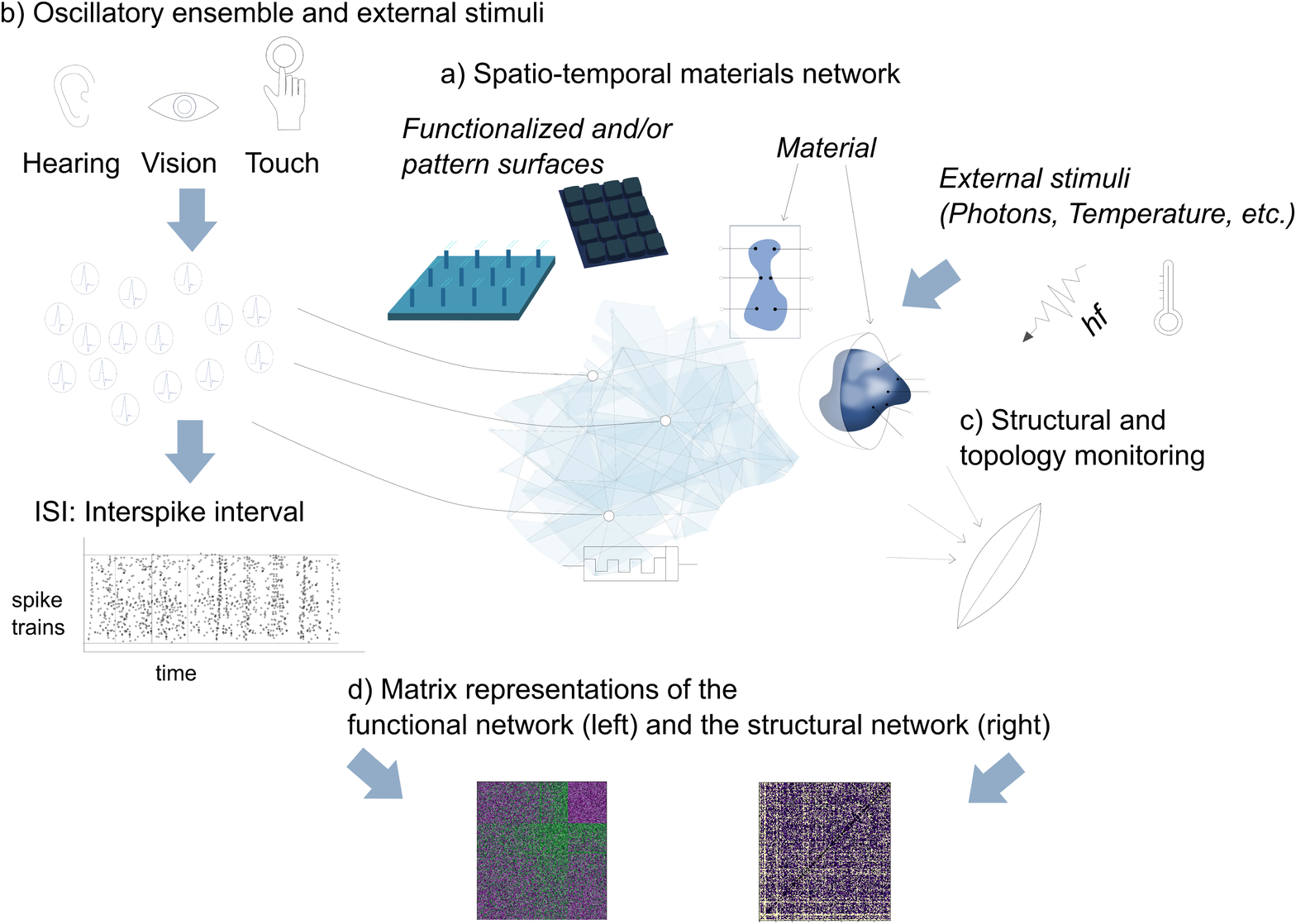}
	\caption{Artificial spatio-temporal networks: Materials science considerations and design strategies to generate higher brain function in artificial systems. The proposed system takes basal functionalities of bio-inspired information pathways into account discussed in Chapters 3 and 4. (a) 2D or 3D spatio-temporal materials network. Wires within the network are connected via memristive components. The memristive functionality at cross-sections of the network implements memory and local plasticity in the network.  The faded area represents a growing network. In the case of a 2D network structure formed on a planar substrate, network growth might be modified, for example, by pre-pattern substrates, a functionalized surface, additional electrical potentials, optical stimuli, and deposition-related growth parameters (materials, deposition rate, reactive gases, substrate temperature, and so on). A 3D network allows further freedom of design and allows for a nervous system-like connectivity. The 3D network could be in a solid phase, or even multiphase, network, the latter combining materials in the solid, liquid, and gas phase. (b) Representation of a pulse-oscillator ensemble in order to mimic neural spiking activity. The individual oscillators of the ensemble are electrically connected to the network, leading to modifications of the network connectivity by oscillator pulses. Conversely, the network weights in turn influence the dynamic state of the oscillator ensemble via pulse coupling.  The oscillatory ensemble allows an input of external stimuli (e.g., touch, vision, and hearing) via fire rate coding. In addition, analyzing the interspike interval (ISI) distributions of the ensemble in quasi-real-time enables permanent monitoring of the dynamic state of individual oscillators, as well as the entire ensemble. (c) Stage to monitor the structure and extract the topology of the network in real time by, for example, optical microscopy, electron microscopy, thermal imaging, or magnetic field distribution detection (similar to MEG (magnetoencephalography)).(d) By monitoring the oscillatory ensemble dynamics (see (b)) and the structural connectivity (see (c)), the spatio-temporal state and its evolution can be analyzed in real-time
}
	\label{fig:fig11}
\end{figure}

Before describing the interplay between the components sketched in Fig. 11, we should first consider a few aspects of biological information pathways which are obviously implementable by materials science and electronics, and might simplify the execution of the proposed artificial spatio-temporal network. 
In a human brain, the ability to access, and thus measure, the structural, topological, and dynamical states is hindered by both technological and ethical constraints \citep{fukushiEthicalChallengesClinical2008, opitzLimitationsExVivo2017}. In contrast, artificial systems should theoretically permit access to all local and global parameters in any conceivable experimental setup. This offers a high degree of freedom in designing artificial systems. In particular, for a system growing in complexity, a designer might decide which segments should be externally controlled and which should develop via self-assembly and self-organization.

Furthermore, the time scales involved in biological information processing may actually facilitate their artificial engineering. In phylogenetic and ontogenetic development, low time scales dominate the scene. Species vary from one generation to the next, with networks growing from days to years. Additionally, nervous system dynamics are in the 100 Hz range, with low transmission velocities on the order of m/s, i.e. the speed of spikes along axons are common. As such, there is no need to build ultrafast artificial systems in order to imitate basal biological information pathway.  Indeed, the deposition or synthesis of any material, e.g. nanoparticles or NWNs, is a growing materials network (Fig. 11 (a)), and can be adjusted to low time scales. In addition, low time scales adapt well in many ways to materials transport parameters, including ionic drift, diffusion currents, and mass transport in general. Biological time scales are easily accessible by electronics, facilitating circuit design, and permitting real-time observation of spatio-temporal system development (Figs. 11 (b), (c), (d)). For example, leaky-integrated-firing of a biological neuron can be technically realized by van der Pol (vdP) oscillators \citep{vanderpolLXXXVIIIRelaxationoscillations1926, pikovskijSynchronizationUniversalConcept2003, ignatovSynchronizationTwoMemristively2016}, compact devices based on VO$_2$ or NbO$_x$, which exhibit a negative differential resistance (NDR) I-V curve \citep{d.leeNbO2BasedFrequencyStorable2018, maffezzoniModelingSimulationVanadium2015, driscollCurrentOscillationsVanadium2012, luoVanPolOscillator2022}, or integrated, mixed-signal circuits \citep{x.chengCMOSIntegratedLowPower2021}. In general, low time scales known form biological information pathways, including external stimuli that affect them, offer an exploration space attainable by materials-related phenomena, electronics, and parameter monitoring.

How can an artificial spatio-temporal computing system, as sketched in Fig. 11, be practically realized?  The goal in a bio-inspired artificial spatiotemporal network is to reach the desired topological and dynamical states simultaneously, in order to mimic the previously discussed characteristic hallmarks of the nervous system. This is handily illustrated in both Fig. 6 (a), where the topological cortical map region is labeled “G” (Goal) in the network cube, and in Fig. 6 (b), where the state of SOC is highlighted as the envisaged dynamical state. The main challenge here is to define the appropriate material network properties and dynamical setting for the entire system that will enable a similar spatiotemporal state to that of a nervous system. This global system state is often said to be structurally complex while being temporally close to the edge of chaos \citep{skardaHowBrainsMake1987, kingFractalChaoticDynamics1991, chialvoEmergentComplexNeural2010, strogatzNonlinearDynamicsChaos2015, mackeyOscillationChaosPhysiological1977, bullmoreComplexBrainNetworks2009, schroeterEmergenceRichClubTopology2015}. To achieve this goal, we describe the components presented schematically in Fig. 11 and their interactions in accordance to the biological information pathways described in Chapters 3 and 4. 
The material network template offers manifold opportunities on either a 2D or 3D platform (Fig. 11 (a)). A network growth mimicking ontogenesis can be realized by continuous film deposition, or ongoing material synthesis of, for example, nanoparticle or nanowire networks. Network growth can be influenced in at least in three ways, the first of which being the oscillatory ensemble that is electrically connected to the network. Here, external stimuli, e.g., hearing, vision, and touch, are imprinted into the material network growth process via fire rate coding (Fig. 11(b)). Network formation and structure evolution are modified by the additional potential differences between the oscillator contacts within the network. Second, by integrating additional conductive pads (islands) on a 2D substrate platform, the formation of filaments between the oscillator’s electrodes can be controlled by the islands’ shape, number, size, and/or additional applied bias potential (Fig. 11(a)). The formation of conductive filaments during network growth could also be manipulated via structurally modulated or functionalized surfaces. In this way, not all network pathways are allowed, while others are assisted \citep{d.michaelisSelfOrganizingGaitPattern2021}. Biologically, this corresponds to axon growth and guidance \citep{hiesingerBrainWiringComposite2021}. This approach can also apply to 3D structures, which provides an increased degree of freedom and in principle allows nervous system-like connectivity.  The materials network, whether 2D or 3D, does not necessarily have to be in the solid state: Electrolytes may be an appropriate fluid which satisfies the aforementioned requirements, including the state of criticality \citep{fisherStoryCoulombicCritiality1994, robinModelingEmergentMemory2021, pantoneMemristiveNanowiresExhibit2018, aokiSelforganizationComplexNetworks2015, hauglandSelforganizedAlternatingChimera2015, patzauerSelfOrganizedMultifrequencyClusters2021, orlikGeneralPrinciplesSelforganization2012, orlikSpatiotemporalPatternsControl2012}. Third, additional stimuli (Fig. 11 (a)) in the form of, e.g.  light or temperature, can also modify the spatio-temporal evolution of the functional material network. An imprint of information during network growth is common to all three methods. This distinguishes the artificial spatio-temporal network approach from common AI systems. In the latter, the training or learning sequence is applied after system manufacture. By applying the three methods described above, it might be possible to imprint information in a similar way to that of a human nervous system during ontogenesis (method 1), as well a kind of a-priori knowledge (methods 2 and 3), i.e. phylogenetic factors.  

For the entire system, a simultaneous, in-depth monitoring of network structure during its development and temporal evolution is intended, in accordance with a neuroscience approach to extract the structure and dynamics of complex brain networks \citep{spornsStructureFunctionComplex2013, bullmoreComplexBrainNetworks2009}. To this end, the spatio-temporal development of time-varying connectivity within the functional materials network (see Fig. 11 (a)) will be monitored, for example, by means by optical microscopy, electron microscopy, or the magnetic field distribution in accordance with magnetoencephalography (MEG) (see Fig. 11(c)). This will allow visualization of the time-evolving correlation matrix of the oscillatory ensemble, and the extraction of more theoretical metrics, such as cluster coefficients, characteristic path lengths, motifs, modularity, and hubs \citep{schroeterEmergenceRichClubTopology2015, bullmoreComplexBrainNetworks2009, fornitoFundamentalsBrainNetwork2016, spornsNetworksBrain2011}. In Fig. 6(a), the pale blue dashed line in the cube represents a fictional pathway through the network cube. At first, we assume that the materials network is a topology state labeled “S” (Start). The position “S” within the cube is chosen as an example, but could just as well be any other topological position within the network cube. By constantly monitoring the topology of the system during network growth and intervention via a set of parameters (e.g. added materials, extra potentials, and external stimuli to the oscillatory ensemble), it might be possible to adjust the system to arrive at “G”, defined by a set of characteristic theoretical parameter (hubs, motifs, modularity, cluster coefficient, path length, etc.) \citep{fornitoFundamentalsBrainNetwork2016}.

Simultaneously, the ISI distribution and time series of the oscillatory ensemble will be recorded (see Fig. 11 (b)) \citep{kreuzMeasuringSynchronizationCoupled2007}. Spike train distances provide a means of quantifying neuronal variability and the degree of synchrony in and between oscillatory ensembles, and may indicate the rise of oscillatory avalanche firing as one indicator of the SOC  \citep{abbottSimpleGrowthModel2007, beggsNeuronalAvalanchesNeocortical2003, scarpettaNeuralAvalanchesCritical2013, priesemannSpikeAvalanchesVivo2014, timmeCriticalityMaximizesComplexity2016, miltonNeuronalAvalanchesEpileptic2012, disantoLandauGinzburgTheory2018}. SOC is described as a state located somewhere between the random, independent firing of individual oscillators, and complete synchrony, where all oscillators fire in phase with the same frequency \citep{bottani1995pulse, aokiSelforganizationComplexNetworks2015}. Between these two extremes, a system’s dynamics can be characterized by a complex and time-varying interaction of the oscillatory ensemble (see Fig. 6 (b)). This regime exhibits features of criticality typically observed close to phase transitions \citep{chialvoEmergentComplexNeural2010, shewInformationCapacityTransmission2011, beggs2012being, srinivasa2015criticality, chialvo2020controlling, mallinsonAvalanchesCriticalitySelforganized2019, pikeAtomicScaleDynamics2020}. In particular, the topology and temporal dynamics of a system in such a state are extremely sensitive to external distortions (stimuli) and may respond to them in numerous ways.

Practically speaking, we will begin by analyzing coupled nonlinear oscillator network raster plots, phase portraits, phase response curves, bifurcation diagrams, spike distance measurements, and cross-correlation type time-series analysis. Information from these analyses will be subsequently applied to quantify the phase and frequency relationships between network oscillators and their development over time \citep{kreuz2009measuring, hoppensteadt2012modeling}.
Finally, we would like to discuss obvious obstacles and challenges. In section 4 D, the rule of homeostasis was highlighted. The concept of homeostasis is of essential importance to stabilize the nervous system dynamics and morphology to a set-point. For the system presented in Fig. 11, homeostasis is not illustrated. It might be possible to reconstruct a feedback parameter from the structural and functional matrices to reduce or enhance, if necessary, the oscillatory activity, or to modify the material growth process. Another challenge might be the implementation of appropriate delay lines to mimic the important signal retardation known from nervous systems \citep{amilOrganizationIdentificationSolutions2015, mackeyOscillationChaosPhysiological1977}. Ionic conductors with slow ionic motion in the form of drift or diffusion currents could be a possible solution. 

One important issue remains: Picture a fabricated artificial spatio-temporal system as depicted in Fig. 11, that presents all previously discussed biological information pathways. How can we benchmark the system, and determine how it solves tasks set by external stimuli? Certainly, the functional and structural network states reflect the overall system state. As such, one viable approach is to read out the system state and to activate a set of artificial motor neurons to react to an input task. However, this does not accurately represent the process in the human brain, where there is no internal, global system observer to decide on the next step \citep{damasioDescartesErrorEmotion2004}. At this point, we are confronted with a difficult challenge: how can we lead matter to imitate the mind? While the authors can suggest an example system as shown in Fig. 11, this question remains open.

\section{Benchmarking for Bio-inspired Computing}

Benchmarking in AI is an important approach to measure its performance, and subsequently enable comparisons between different systems. In pattern recognition, for example, MNIS data sets are used, with the recognition rate defining a clear benchmark. While contemporary AI systems show extraordinary capability in performing a single, specific task, their success at task variability is highly limited compared to the nervous system. Nonetheless, a new generation of AI has demonstrated extraordinary capabilities in the field of gaming (Chess and Go), including an aptitude for self-learning \citep{silverGeneralReinforcementLearning2018}. Yet, it remains unclear how to define a fair and comprehensible benchmarking for neuromorphic systems and bio-inspired computing \citep{daviesBenchmarksProgressNeuromorphic2019}. Computational tasks must be carefully designed in order to assess the overall system’s performance in comparison with human mental capabilities, as previously proposed by Alan Turing in his seminal work on Machinery and Intelligence \citep{turingCOMPUTINGMACHINERYINTELLIGENCE1950}. Bloom’s learning taxonomy, which was developed to hierarchically categorize learning in the classroom, can be helpful in assessing how successfully artificial systems mimic higher brain functions \citep{adamsBloomTaxonomyCognitive2015}. This taxonomy contains six categories of cognitive skills and presents a hierarchy with increasing cognitive functionality from bottom (factual knowledge) to top (creation) (see Fig. 12), or in other words, from lower-order skills that require less cognitive processing to higher-order skills that require deeper learning and a greater degree of cognitive processing \citep{compeauEstablishingComputationalBiology2019}. 

\begin{figure} [ht!]
	\centering
	\includegraphics[width=0.7\textwidth]{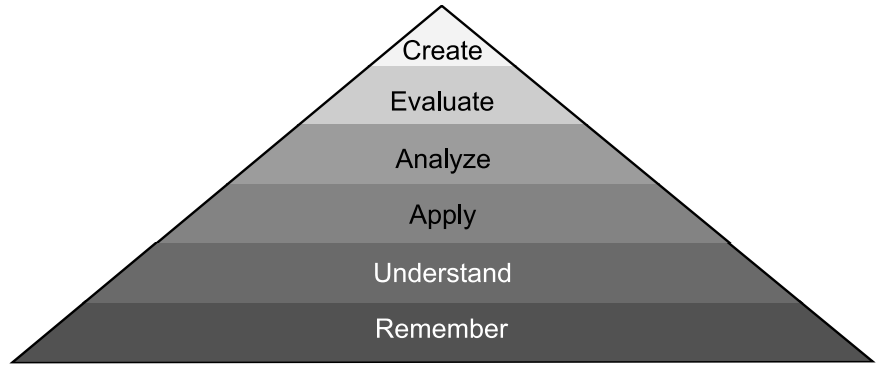}
	\caption{Suggested benchmark for bio-inspired systems based on Bloom’s taxonomy. The pyramid represents increasing cognitive human skills from bottom to top. Fig. 2 from Ref. \citep{compeauEstablishingComputationalBiology2019}.}
	\label{fig:benchmark}
\end{figure}

This strategy may serve as a basis for benchmarking in bio-inspired computing systems. However, due to the nervous system’s task variability for each of the six cognitive categories, transparent benchmarks must be developed. This goal is extremely important for future comparisons of bio-inspired systems, which are currently developed on different platforms. In addition, resource-related parameters, such as energy consumption, system weight, and failure tolerance, need to be included.

\section{Discussion}

This perspective introduces the concept of artificial spatio-temporal networks, which proposes basal hallmarks, such as morphological and dynamical characteristics of nervous systems, to reproduce higher brain functions in artificial systems. In particular, the basal mechanisms known from the growth of nervous systems might play a significant role in their function. This concept will undoubtedly be a way to include biologically-relevant features in future artificial systems. Yet, only the tip of the iceberg has thus far been addressed: to fully realize an artificial spatio-temporal network, several challenges remain unresolved.

In more general terms, artificial spatio-temporal networks again raise the fundamental question: “To what extent can higher brain functions be reproduced in artificial systems?” Seminal books and papers by \citep{churchlandCouldMachineThink1990, churchlandEngineReasonSeat1995, aleksanderHowBuildMind2001, kochCanMaschinesBe2008, dehaeneWhatConsciousnessCould2017, hiesingerSelfassemblingBrainHow2021, melloniMakingHardProblem2021}, and many more address this topic in one way or another. According to the authors, higher brain function can be described on the basis of natural sciences and mathematics, permitting us to view this challenge in another light. On an atomistic level, we find in living nature, and therefore in any nervous system, old and well known friends from the periodic table of the elements, including but not limited to Carbon (C), Sodium (Na), Potassium (K), Chlorine (Cl), Oxygen (O), and Hydrogen (H). Any effort to establish higher brain function in an artificial system, whether in silico (software oriented) or in a material-based substrate, as in the case of artificial spatio-temporal networks, should apply another tool box of elements to establish awareness, perception, or consciousness, e.g. Silicon(Si), Gold (Au), Silver (Ag), Tungsten (W), O, and so on. There is no obvious reason why this strategy should not work, but if it cannot, what are the fundamental limits, and how are they defined? A look at biochemical substrates in living species highlights the weaknesses of the simplistic, atomistic view point. There is still unknown genetic information that strongly controls nervous system behavior and function, especially during development, which therefore cannot currently be considered in any artificially-constructed systems. Whether there are shortcuts to bypass the role genes play in neural behavior and development is completely unknown, and might act as a show stopper \citep{hiesingerSelfassemblingBrainHow2021}. On the one hand, it is truly challenging to introduce basal biological functionalities, such as homeostasis, signal delay, growth, and the appropriate states of criticality and topology in an artificial system. On the other hand, the materials tool box may offer plethora of phenomena which have not yet been explored for novel computing architectures \citep{kasparRiseIntelligentMatter2021, orlikGeneralPrinciplesSelforganization2012, orlikSpatiotemporalPatternsControl2012, hauglandSelforganizedAlternatingChimera2015, patzauerSelfOrganizedMultifrequencyClusters2021}.  Hence, these simple questions and views point towards an even more fundamental aspect: in living systems,  the separation between matter and information becomes blurred, making it risky to apply these terms without investigating living and artificial systems equally, or precisely clarifying the respective context \citep{johannsenInformationUndIhre2016}.    

\section{Conclusion}

In this perspective, we addressed fundamental limits of current ICT and briefly summarized the state-of-the-art. Today’s digital electronics work with clock rates in the GHz range, utilizing ns pulses and signal transmissions at nearly light speed in a vacuum. Meanwhile, nervous systems exhibit numerous remarkable and fascinating features, including anticipation, awareness, perception, and consciousness. The associated action potential spikes are 6 orders of magnitude longer, and travel with a velocity 6 orders of magnitude lower, than their electronic analogs, while dissipating only a couple of Watts of power. We touched on the fundamentals of information processing in biological (nervous) and engineered systems. Specifically, we highlighted the dynamical and morphological properties exhibited by nervous systems using the human brain as an example. The exceptional topology of the human cortex in comparison to other biological and technical networks, in addition to the state of SOC, served as guidelines to develop artificial spatio-temporal systems. A pathway to realize artificial spatio-temporal systems in a hardware-orientated system was presented, aiming to emulate higher brain functions in an artificial system. The role of ontogenesis was discussed, revealing that the mechanism of neural network growth provides crucial information useful in designing novel artificial computing systems, which have yet to be addressed in great detail. 

Neural network growth illustrates how important the ongoing interaction between the internal and external world is when artificially creating the basic structures that provide the ability to learn specific functions. In our opinion, this emphasizes the importance of basal properties which, while beginning to be applied systems in artificial, have yet to be fully implemented. These properties include individual autonomous dynamic units, time-variable coupling between them, and both positive and negative connection growth. With respect to time-variability, the research field has shown enormous progress in recent years with the development of memristive systems. Although memristive devices can already replicate the phenomena associated with learning to a certain degree, the question remains whether these devices can suitably reproduce both the necessary processes in their entirety, and global dynamics which are shaped by an overwhelmingly complex network. The last point in particular presents immense challenges for a conservative implementation of memristive devices in large-scale systems. Finally, we discussed possible limitations in implementing higher brain functions in artificial systems. We concluded that genetic information plays a key role in the development of neural nervous systems, knowledge that we are still lacking if we want to fully implement this behavior in artificial systems, specifically with regards to awareness, perception, and consciousness. The exploration space for implementation is certainly extraordinary large for artificial spatio-temporal systems. This huge parameter space is both curse and blessing: while such a large number of variables must be monitored and controlled, it also allows for greater flexibility and opportunities.  
One thing is certain in this context: no matter which engineered solution ultimately prevails, humanity will be confronted with a multitude of ambivalent questions and challenges, in which certainly “Matter \& Mind Matter”.

\section*{Acknowledgments}

We thank Nora Kohlstedt for preparing part of the figures. “Funded by the Deutsche Forschungsgemeinschaft (DFG, German Research Foundation) – Project-ID 434434223 – SFB 1461". The project is entitled SFB 1461 “Neurotronics: Bio-inspired Information Pathways” (see for more details https://www.crc1461-neurotronics.de/index.php/en/). Moreover, the financial support by the DFG via the Research Unit 2093: “Memristive Devices for neural Systems” is acknowledged. We thank Gitanjali Kolhatkar and Shane Scott for carefully reading the manuscript.

\bibliographystyle{plainnat}  
\bibliography{references}

\end{document}